\documentclass[a4paper,fleqn,usenatbib]{mnras}

\usepackage{newtxtext,newtxmath}
\usepackage{tikz,xcolor,hyperref}
\definecolor{lime}{HTML}{A6CE39}
\DeclareRobustCommand{\orcidicon}{%
	\begin{tikzpicture}
	\draw[lime, fill=lime] (0,0) 
	circle [radius=0.16] 
	node[white] {{\fontfamily{qag}\selectfont \tiny ID}};
	\draw[white, fill=white] (-0.0625,0.095) 
	circle [radius=0.007];
	\end{tikzpicture}
	\hspace{-2mm}
}

\foreach \x in {A, ..., Z}{%
	\expandafter\xdef\csname orcid\x\endcsname{\noexpand\href{https://orcid.org/\csname orcidauthor\x\endcsname}{\noexpand\orcidicon}}
}

\usepackage{newtxtext,newtxmath}

\usepackage[T1]{fontenc}
\usepackage{ae,aecompl}

\usepackage{graphicx}	
\usepackage{amsmath}	
\usepackage{xspace}

\title[Filament extraction in GAEA]{The filament determination depends on the tracer: comparing filaments based on dark matter particles and galaxies in the GAEA semi-analytic model} 

\author[Zakharova et al.]{Daria Zakharova$^{1,2}$\orcidA{}\thanks{E-mail: dzakharovaa@gmail.com}, Benedetta Vulcani$^{2}$\orcidB{}, Gabriella De Lucia$^{3,4}$\orcidC{},  Lizhi Xie$^{5,3}$\orcidD{}, \newauthor 
Michaela
Hirschmann$^{6,3}$\orcidE{}, Fabio Fontanot$^{3,4}$\orcidF{} \protect\\
$^{1}$ Dipartimento di Fisica e Astronomia “G. Galilei”, Universita di Padova, Vicolo dell’Osservatorio 3, I-35122 Padova, Italy
\\
$^{2}$ INAF-Padova Astronomical Observatory, Vicolo dell’Osservatorio 5, I-35122 Padova, Italy
\\
$^{3}$ INAF – Osservatorio Astronomico di Trieste, Via Tiepolo 11, I-34131 Trieste, Italy
\\
$^{4}$ IFPU - Institute for Fundamental Physics of the Universe, via Beirut 2, 34151, Trieste, Italy
\\
$^{5}$ Tianjin Normal University, Binshuixidao 393, 300387, Tianjin, China
\\
$^{6}$ Institute for Physics, Laboratory for Galaxy Evolution and Spectral modelling, Ecole Polytechnique Federale de Lausanne,\\
Observatoire de Sauverny, Chemin Pegasi 51, 1290 Versoix, Switzerland
}

\date{Accepted XXX. Received YYY; in original form ZZZ}

\pubyear{2023}

\begin{document}
\label{firstpage}
\pagerange{\pageref{firstpage}--\pageref{lastpage}}
\maketitle

\begin{abstract}
Filaments are  elongated structures that connect groups and clusters of galaxies and are visually the striking feature in cosmological maps. In the literature, typically filaments are defined only using galaxies, assuming that these are good tracers of the dark matter distribution, despite the fact that galaxies are a biased indicator. Here we apply the topological filament extractor DisPerSE to the predictions of the semi-analytic code GAEA to investigate the correspondence between the properties of $z=0$ filaments extracted using the distribution of dark matter and the distribution of model galaxies evolving within the same large-scale structure. We focus on filaments around massive clusters with a mass comparable to Virgo and Coma, with the intent of investigating the influence of massive systems and their feeding filamentary structure on the physical properties of galaxies. We apply  different methods to compare the properties of filaments based on the different tracers and study how the sample selection impacts the extraction. Overall, filaments extracted using different tracers agree, although they never coincide totally. We also find that the number of filaments ending up in the massive clusters identified using galaxies distribution is typically underestimated with respect to the corresponding dark matter filament extraction.
\end{abstract}



\section{Introduction}
\label{section:intro}
On scales of the order of a few Mpcs, the  distribution of dark matter and galaxies in the Universe is not uniform. These   
inhomogeneities, that are  well visible  in observations~(SDSS~\citealt{Guo+2015}, ~\citealt{Duque+2022}; 2MASS~\citealt{Skrutskie+2006}, \citealt{Huchra+2012}, SAMI~\citealt{Bryant+2015}), are also identified in cosmological simulations assuming a cold dark matter and an accelerated expansion of the Universe ~\citep{Bond+1996, Springel+2005} and define what is often referred to as `cosmic web'. The formation of large-scale structure is predicted as a natural outcome of the evolution of the non-linear growth of primordial density perturbations~\citep{Peebles_1980_book}: over-dense regions grow in size and become denser, while underdense regions expand under the action of gravity~\citep{Scoccimarro+2000}. As a result, the cosmic web can be described as an ensemble of 
haloes of different mass connected by filaments 
\citep{Codis+2018}. 
\par
The formation and evolution of galaxies takes place inside the cosmic web, and the properties of galaxies are inextricably linked to their environment. 
Galaxies in rich clusters tend to have a more spherical shape, to be redder and more massive~\citep{Dressler+1980_red, Boselli+2006},  
to have lower star formation activity ~\citep{Vulcani+2010, Poggianti+2006, Paccagnella+2016}, and contain less gas~\citep{Haynes+1985} than their field counterparts.  Galaxies in filaments have properties that are intermediate between clusters and the field~\citep{Castignani+2022_2, Castignani+2022_1}. They are typically more massive, redder, more gas poor, and have earlier morphology than galaxies in voids~\citep{Rojas+2004, Hoyle+2005, Kreckel+2011, Beygu+2017, Kuutma+2017}, but have later type morphology and are more star forming  than cluster galaxies (e.g. \citealt{Castignani+2022_1, Castignani+2022_2}).
Recent observational works also suggest that they have more extended ionized gas distributions~\citep{Vulcani+2019} and reduced atomic HI gas reservoirs~\citep{Kleiner+2017, Odekon+2018, Blue_Bird+2020,  Lee+2021}. 
Various mechanisms have been proposed as responsible for the observed trends, e.g.
mergers, stripping, tidal interactions, and starvation~\citep{Boselli+2006, De_Lucia+2012, Kraljic+2018, Donnari+2021}. However, 
the relative role of the different physical mechanisms as a function of  environment still needs to be quantified~\citep{Song+2021, Chang+2022}.
\par
Many different tools have been developed to identify the different components of the cosmic web, including algorithms to identify filamentary structure. These include methods based on particle density distribution analysis~(DisPerSE, \citealt{Sousbie+2011}),  on density and tidal fields analysis~(NEXUS+, \citealt{Cautun+2015}), velocity shear tensor-based cosmic web analysis~(COWS, \citealt{Pfeife+2022}), and deep learning~\citep{Inoue+2022}.
A detailed comparison of many of these algorithms has been presented in \citet{Libeskind+2018}. Albeit different filament finders are available,  an operational and rigorous definition of a "filament" is still missing.
Different studies adopt different detailing of the cosmic web and use datasets with different characteristics, with the result that they are sensitive to different properties of the structures to be identified. 
The availability of large spectroscopic samples has provided significant impetus, in the last years, to both theoretical and observational work focusing on the detection and analysis of filamentary structure.
Theoretical studies (e.g., ~\citealt{Aragon-Calvo+2007, Cautun+2014, Chen+2015, Laigle+2018, Kraljic+2019, Kuchner+2020, Kuchner+2021, Rost+2021}) have the advantage to be able to study both dark matter and galaxy distribution. Typically,  the tracer used depends on the aim: dark matter particles are  used to characterize the cosmic web structure (e.g. \citealt{Cautun+2014, Bermejo+2022}), 
while galaxies are used 
when interested in characterizing the effect of environment on galaxies 
(e.g. \citealt{Singh+2020, Galarraga-Espinosa+2020, Kuchner+2021}). 
\par
Observational studies (e.g.,~\citealt{Tempel+2014, Alpaslan+2014,  Laigle+2018, Kraljic+2018, Malavasi+2017, Malavasi+2019, Malavasi+2020, Castignani+2022_1})
do not have access to the dark matter distribution and typically rely on the assumption that
the distribution of galaxies reflects well enough that of the dark matter. This is true only in first approximation: albeit the two components are intrinsically linked, 
the density of galaxies is a non-trivial
function of the  dark matter density
(e.g.~\citealt{Kaiser+1984, Tegmark+1998}), with the result that some galaxies are locally more (``biased'') or less (``anti-biased'') clustered relative to dark matter. 
\par
Very few studies have so far focused  on  the difference between the determination of filaments using galaxies and dark matter, and these have mainly used hydrodinamical simulations.  For example, \cite{Laigle+2018} compared the skeletons obtained from the distribution of galaxies  in  the COSMOS2015~\citep{Laigle+2016} survey, and  from the distribution of galaxies  and dark matter in realistic mock catalogues  extracted from a lightcone built from the cosmological hydrodynamical simulation HORIZON-AGN~\citep{Dubois+2014}.
They found that a (small) fraction of dark matter filaments have no counterpart in the distribution of galaxies.
\par
In this paper, we characterize the cosmic web of simulated cosmological boxes, using separately
dark matter particles and galaxies as tracers
 {in} the semi analytic model GAEA \citep{Xie+2020}. We focus on simulated regions  around massive haloes, with the intent of investigating the influence of massive systems and their feeding filamentary structure on the physical properties of the galaxies.
The two most studied clusters and infalling filaments in the local Universe are Virgo~\citep{Tully+1982, Kim+2016, Castignani+2022_2, Castignani+2022_1} and Coma~\citep{Malavasi+2020}. Hence we focus on the filaments around simulated haloes of similar mass. 
While this paper is devoted to a purely theoretical analysis, in future work we will compare theoretical and observational results, and make dedicated predictions for testing the role of filaments in affecting the galaxy properties. 
\par
The paper is organized as follows. In \hyperref[sec:data_and_methods]{Section 2}, we provide a brief description of the semi analytic model used in our study, and describe the sampling and filament extraction method. In \hyperref[sec3:properties]{Section 3}, we characterize the  filamentary structures identified, and compare filaments detected in dark matter with their counterparts based on the galaxy distribution. In \hyperref[sec:different_settings]{Section 4}, we discuss how galaxies with different properties (stellar mass and galaxy type) track filaments.
We discuss and summarize our results in  \hyperref[sec:discussion]{Section 5}.

\section{Data and methods}
\label{sec:data_and_methods}

\subsection{GAEA}

In this work, we  {use}
the GAlaxy Evolution and Assembly (GAEA) semi-analytic model of galaxy formation and evolution. In particular, we use the model version that has been published in \citet{Xie+2020} that includes: (i) a detailed treatment for the chemical enrichment that accounts for the non instantaneous recycling of gas, metals and energy \citep{DeLucia+2014}, (ii) an updated treatment for the stellar feedback that provides a good agreement with the observed evolution of the galaxy mass-gas metallicity relation and of the galaxy stellar mass function up to $z\sim 3$ \citep{Hirschman+2016}, and (iii) an explicit treatment for partitioning  the cold gas in its atomic and molecular component and for ram-pressure stripping of both the hot gas and cold gas reservoirs of satellite galaxies \citep{Xie+2017,Xie+2020}. The model realization is coupled to the Millennium Simulation \citep{Springel+2005} - a dark-matter only N-body simulation of a box of side length equal to 500 Mpc/h comoving. The simulation adopts a WMAP1 cosmology with $\Omega_{\rm m}=0.25$, $\Omega_{\rm b}=0.045$, $\Omega_{\Lambda}=0.75$, $h=0.73$, and $\sigma_8=0.9$. While more recent determinations suggest a lower value of $\sigma_8$, we do not expect this to have a significant impact on the results of this work other than on the number of massive haloes identified in the simulated box at $z=0$. The particle mass of the simulation is $m_{\rm DM} = 8.6\times10^8 {\rm M}_{\odot}\,h^{-1}$, which translates in a galaxy stellar mass limit of approximately $10^9\,{\rm M}_{\odot}$. In our analysis, we have used galaxies more massive than $10^9\,{\rm M}_{\odot}$. GAEA follows the evolution of dark matter substructures until they disappear (i.e., they are stripped below the resolution of the simulation).   When substructures are very close to the detection limit (20 particles), these could have issues with spuriousness at the low-mass end. However, model results are not significantly affected by this. Previous papers~\citep{Hirschman+2016, Xie+2017} have shown that model results converge down to galaxy stellar masses ~$10^9\,{\rm M}_{\odot}$. 
        
Our study focuses on a detailed comparison between the filamentary structure identified using simulated galaxies and dark matter particles, when the full 3D information is available. In particular, we focus on the simulated volume at $z=0$, and select independent sub-volumes corresponding to boxes of 70 Mpc/h on a side. This size is similar to that typically adopted in recent studies on the environmental influence and properties of galaxies based on state-of-the-art hydrodynamical simulations (50 Mpc/h, TNG50,~\citealt{Park+2022}, or 100 Mpc/h,TNG100~\citealt{Donnari+2021}, EAGLE~\citealt{Singh+2020}, HORIZON AGN~\citealt{Laigle+2017}), and it is sufficient to investigate a wide range of environments, from very sparse regions to very massive and dense structures. 

As mentioned in Sect.~\ref{section:intro}, the most investigated cosmic web regions in the local Universe are the Virgo and Coma clusters, located respectively 16 and 99 Mpc from us and with halo mass of $5 \cdot 10^{14} M_{\odot}$~\citep{Groener+2016} and $6\div9 \cdot 10^{14} M_{\odot} $~\citep{Okabe+2014}, respectively. A number of observational studies investigate the filamentary structure around these clusters \citep[e.g.][]{Castignani+2022_1, Castignani+2022_2, Malavasi+2019}, and future surveys will provide additional data for these regions (4HS~\citep{Taylor+2020}, DESI~ \citep{DESI+2016}).
To mimic the region around the Coma cluster, we focus on haloes with mass  $6.0 \cdot 10^{14} M_{\odot} \le M_{h} \le 1.8 \cdot 10^{15} M_{\odot}$ from the simulated volume. We also require that the selected haloes have no other companion haloes inside each cube 
with a mass equivalent to or larger than Virgo, to mimic real Virgo or Coma clusters.
Only 9 haloes meet these criteria, within Millenium Simulation, and we will consider all of them in the analysis presented below.
Similarly, to reproduce the Virgo cluster, we select haloes with $4.7 \cdot 10^{14} M_{\odot} \le M_{h} \le 6.0 \cdot 10^{14} M_{\odot}$, and require them to be isolated and have no other Virgo-like haloes or more massive ones within the box. We find 12 haloes meeting our selection criteria, and we randomly extract 9 of them to be used in the following. Finally, as a comparison sample, we also consider 9 cubes centered randomly in the simulation to mimic the general field. Therefore, our analysis is based on 27 simulated sub-volumes that are representative of three different large-scale overdensity environments. 

    \subsection{Filament extraction (DisPerSE)}
    \label{sec:disperse_settings}
        To identify filaments, we use the topological filament extractor DisPerSE~\citep{Sousbie+2011}, a commonly used tool to characterize the large-scale structure  distribution~\citep[e.g.][]{Dubois+2014, Bonjean+2020, Barsanti+2022}.
        It identifies persistent topological features in two steps: first,  DisPerSE evaluates the density distribution of tracers using a Delaunay tessellation algorithm~\citep{Weygaert+2009} from an input of discrete positions of tracers (either in 3D or 2D). In a second step, DisPerSE calculates density distribution gradients and identifies zero gradient points (critical points such as minima, maxima and saddle points), as well as all the  segments that trace the ridges of the density field. In this way, the  "skeleton" of the distribution  is constructed, and DisPerSE gives as output information about the filament structure (FS) as a set of critical points (also called vertices) and lines (defined by points) connecting them. The lines between the vertices are also called ``segments'' of the skeleton.
        
        To assess the robustness of the filamentary network and control the scales at which filaments are found, DisPerSE allows the user to tune a signal-to-noise criterion, called {\it persistence}: 
        a filament is identified as an integral line connecting two critical points that represent a critical pair. The persistence of such a pair is the difference (or ratio) of the density values at these points,
        and can be expressed in terms of standard deviations ($\sigma$) of a minimal signal-to-noise ratio or of a cut off value \citep[see][for further details]{Sousbie+2011, Sousbie_etal+2011}. Depending on the goal of the analysis,  it is therefore possible to decide whether faint tendrils should be included (with a trade-off of increased noise) or if the analysis should focus on large scale, collapsed cosmic filaments corresponding to a large signal-to-noise ratio. In this analysis, we are interested in the filamentary structure around massive haloes, and we will favour large values of persistence~ {(or threshold level)} to identify only the predominant structures, while losing detail on small scales.

        Another important parameter that can be tuned while running DiSPerSE is the "smoothness" which reduce the integral lines 
        inhomogeneities. 
        It is  possible~( {but not necessary}) to apply a smoothing procedure  {N times} that averages the position of  {N} point {s} of the filament.  {This setting will smooth out the lines of the skeleton of the filaments.}
        While a larger value of the smoothing parameter might define the filament structures better, smoothing too much leads to a shift of the filament axis and will affect, for example, the density profiles. So the value of this parameter should chosen with care. The level of smoothing also affects the final length of the skeleton.  {The smoothing parameters selected in this work are indicated below (we have verified that the adopted smoothing induces no shift in the density profile). We stress again that  the smoothing procedure is not necessary and does not affect the results of this work.}

        In this work, we are interested in investigating the difference between the filaments based on galaxies and those based on dark matter particles. Therefore, we run DisPerSE on each extracted cube twice, once using the positions of galaxies as predicted by the GAEA model and once using the positions of the dark matter particles from the snapshots of the simulation. In both cases, we use the 3D positions of the tracers. 
        From now on, we will refer to the network obtained in the first run as ``Galaxy filament structure`` (GFS), and to the  the network obtained in the second run as ``Dark matter filament structure`` (DMFS). Since the number of galaxies and that of dark matter particles are very different in the cube considered (see below), adopting the same parameters to define the filament structures would entail the derivation of very different networks, complicating the comparison. To avoid this, we set the parameters for the GFS to best reproduce the visually observed filament structure, and then fine tune the parameters for the DMFS so that the total length of the identified filaments network is comparable to that of the GFS. 
        
        In the case of GFS, only  galaxies with $M_{*} > 10^{9} M_{\odot}$ are considered, as  this is the resolution limit of the model applied to the Millennium Simulation. Each  cube 
        comprises from 10 to 40 thousands galaxies. This variation is related to the different number of haloes with $M_{h} > 10^{14} M_{\odot}$ in each cube and to  their  mass. 
        Regardless of the different number of objects in the different cubes, we use the same values of the DisPerSE parameters for all the 27 cubes, and set the persistence threshold level to  {$10^{4}$}.   {This persistence level best reproduces the visually observed filament structure. We adopt the same value for all the 27 cubes, having checked that the cube-by-cube variation is not significant.}
        This is a relatively high cutoff value, so that only the topologically most robust filaments are included in the analysis.  
        \par
        To further reduce the level  of noise and 
        inhomogeneities, we apply to our skeleton the  DiSPerSE smoothing procedure  5 times,  i.e., averaging
the positions of 5 adjacent segments forming the filament. Again, this value is chosen after visual inspection. In addition, we removed the 10 percent of the  shortest~(by number of points)\footnote{These filaments are also the shortest in length.} filaments.
        \par
        The total length of the obtained skeleton based on galaxy distribution,  averaged over the 27 cubes,  
        is  $L_{GFS} = 840 \pm 162$ Mpc/h. The 
    characteristic length, defined as 
    the total skeleton length divided by the total cube volume  is $0.0024 \pm 0.0005$ Mpc/Mpc$^3$. This value is in good agreement with what found by other studies (e.g. \citealt{Malavasi+2017, Galarraga-Espinosa+2020}). 
            \par
    We now use the total length of the GFS to set the parameters to extract the DMFS.  {We run DisPerSE using all the DM particles}. In this case, each cube contains from 2 to 4 $\cdot 10^{7}$ particles. To obtain a total length that is comparable, within the errors, to that of the GFS, we find that we need to increase the cut off threshold to  {$10^{9}$} and set the smoothness parameter to 3000. These much higher values with respect to those adopted for the GFS are due to the fact that DM particles are at least a factor of thousand more numerous than galaxies. 
    Finally, we remove the 10 percent of the shortest filaments in each cube to get rid of small structures that would only add noise to the analysis. The average total length of the DMFS skeletons  is $L_{DMFS} = 847 \pm 179$ Mpc/h (corresponding to a characteristic length of  $0.0025 \pm 0.0005$ Mpc/Mpc$^3$). 
    \par
    Since the number of particles in the dark matter cube is 
    about 30 times larger than the number of galaxies, 
    also the corresponding output skeleton  {DMFS} contains about 30 times more points than  {skeleton GFS}~(there is a higher density of reference points along the lines describing the filaments). As a result, DMFS are specified with a higher sampling than GFS. We hence reduce the number of DMFS points, preserving the shape of the skeleton,  {so that the median segment of the DMFS is roughly equal in length to the median segment of the GFS}.  This procedure is necessary to perform meaningful comparisons of the distance of galaxies/dark matter particles from the skeleton, as done in the next section. 
        
     \begin{figure*}
        \centering
         \includegraphics[width=1\linewidth]{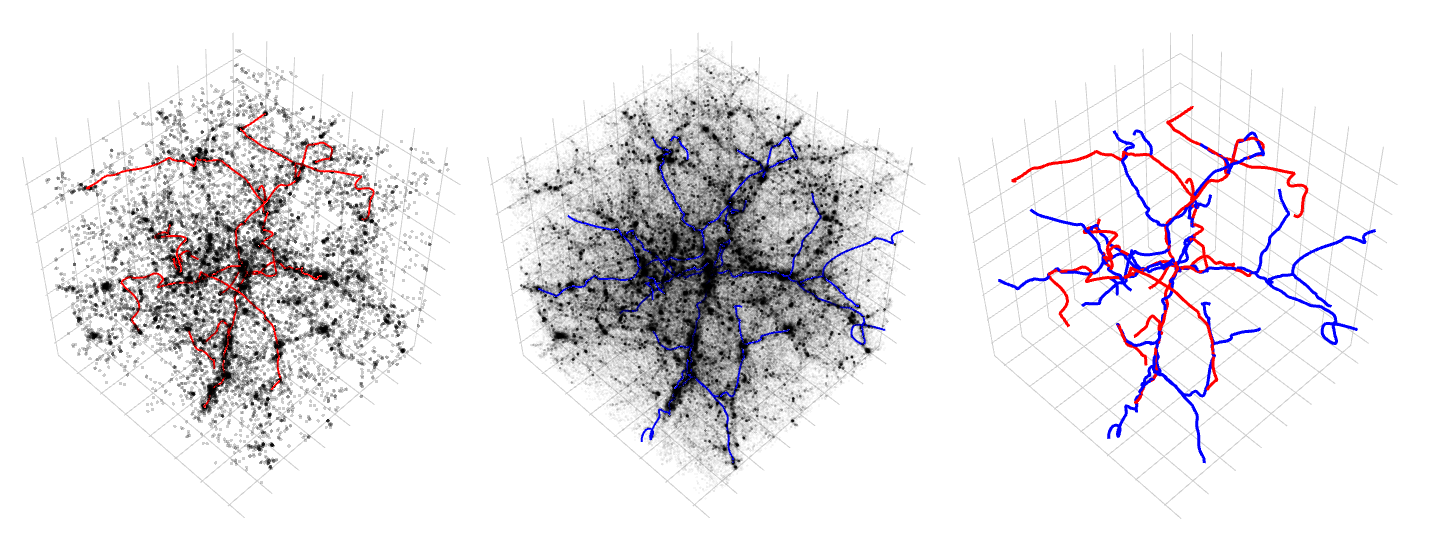}
         \caption{ {Example of the filamentary structures obtained using DisPerSE. The left panel shows the distribution of all galaxies with $M_{\ast} > 10^9 M_{\odot}$ (black dots ) and the filament structure extracted using these galaxies as tracers (red  {curves}). The middle panel shows the distribution of the dark matter particles~ in the same sub-volume of the Millennium simulation. For the sake of clarity, we only plot 2\% of all particles, even though all particles are considered when running DisPerSE. The blue  {curves} show the extracted filaments. The right panel compares the filamentary structures based on galaxies (red) and dark matter (blue). Each cube is 70 Mpc/h on a side.}}
        \label{fig:DisPerSE_res}
    \end{figure*}
       
        Fig.~\ref{fig:DisPerSE_res} shows an example of the final network obtained by DisPerSE for one of our cubes with a Coma-like halo in the center.  The left panel shows the galaxy distribution and the corresponding GFS~(red  {curves}), the middle panel shows the DM distribution and the corresponding  DMFS (blue  {curves}), and the right panel illustrates the two filament structures, for an easy visual comparison. While  
        the two networks are broadly consistent to first approximation, and this is expected as  baryonic matter follows approximately the dark matter, some discrepancies are already evident.

\section{Dark matter and galaxy filamentary structure}
\label{sec3:properties}
    We now aim at quantifying the similarities and differences between GFS and DMFS, to help understanding 
    results from different works, and guide the interpretation of observational data, for which we do not have access to the  information about the dark matter distribution. 
    \subsection{Filament length}
        As already discussed, the definition of what a single filament is varies greatly depending on the approach used to define the skeleton \citep{Kuchner+2021, Galarraga-Espinosa+2020}. In this work, we define filaments as DisPerSE integral lines that connect maxima points~(i.e. the nodes of the skeleton) with saddle-points. 
        \par
        Although -- by construction --  the total length of the skeleton is comparable for DMFS and GFS, some differences in the skeleton properties are observed (Fig.~\ref{fig:DisPerSE_res}). First of all, skeletons extracted by different tracers consist of different numbers of filaments. 
        For all but two cubes, the number of  filaments in the GFS exceeds the number of filaments in the DMFS. The median difference between number of filaments in the DMFS and GFS, considering all cubes is 4, but can be as large as 13.
        \par
\begin{figure}
                \centering
                \includegraphics[width=1\linewidth]{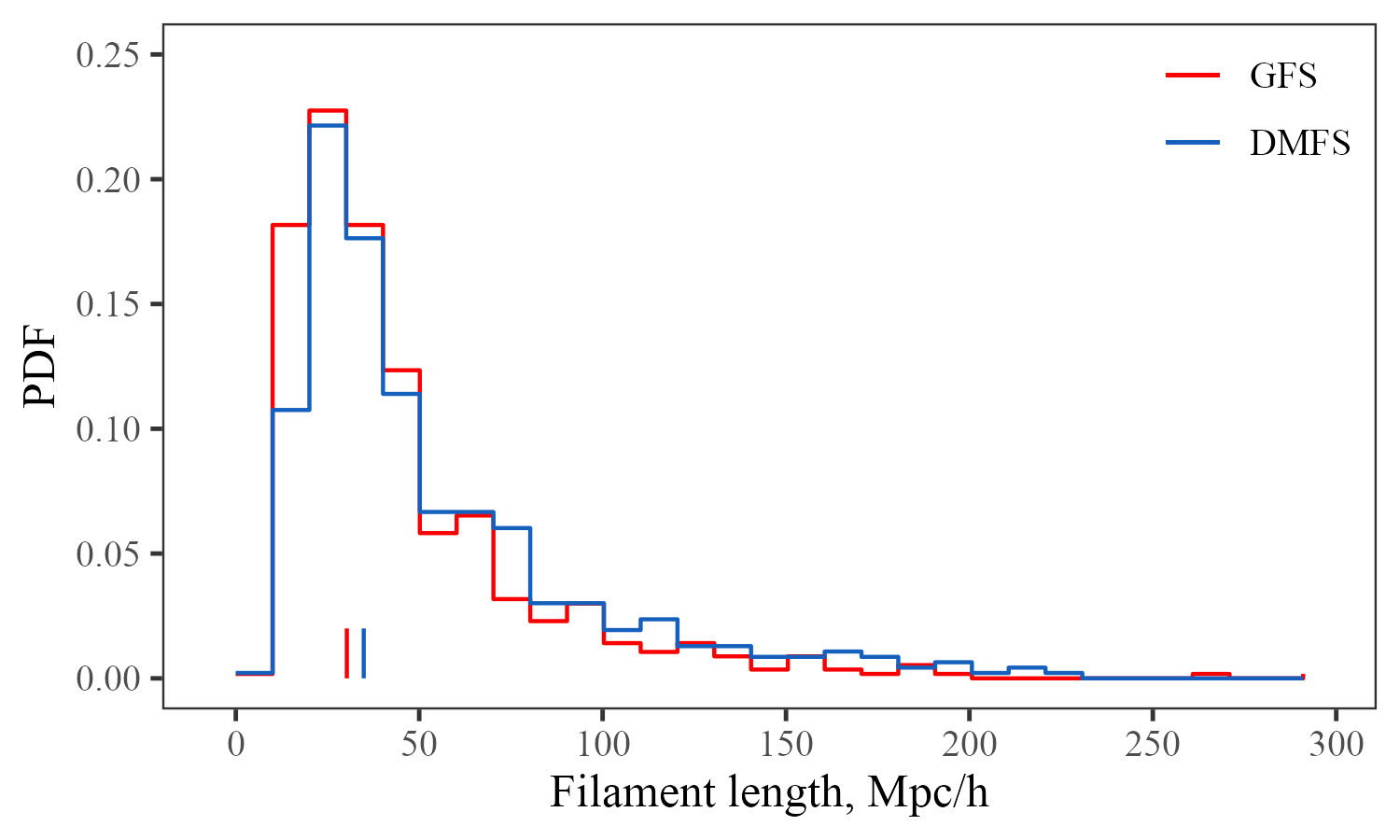}
                \caption{Probability  distribution function (PDF) of the length of each type of filaments (GFS: red; DMFS: blue) for all cubes considered. The small vertical segments indicate the median length of the DMFS ($\sim$35 Mpc/h) and GFS ($\sim$30 Mpc/h).}
                \label{fig:fil_len}
            \end{figure}

        Obviously, if the total length in the GFS and DMFS is the same, and the number of filaments is larger in the GFS, 
        dark matter filaments must be longer. The filaments length distribution is shown in Fig.~\ref{fig:fil_len}. The median length of DMFS filaments is 35$\pm$5 Mpc/h versus 30$\pm$5 Mpc/h for GFS, indicating that indeed overall  DMFS filaments are  always longer than those defined by galaxies (we have run a KS test on the distributions shown in Fig.~\ref{fig:fil_len}, and found a p-value of 0.01468). The length of a filament represents the characteristic distance between critical points for the sampled cube. So, different values of filament length in the DMFS and GFS reflect topological differences between the distribution of dark matter and galaxies with $M_{\ast} > 10^{9}$. Moreover, \cite{Galarraga-Espinosa+2020} notes significant differences for short and long filaments\footnote{$L_f < 9$ Mpc/h for short and $L_f \ge 20$ Mpc/h for long filaments for the volume and DisPerSE settings used in \cite{Galarraga-Espinosa+2020}.}, which also indicates that the length of the filaments is related to the properties of the distribution~(topology) of the particles.
        Finally, we note that unlike the number of filaments, 
        the length of each filament depends strongly
        on the smoothing of the skeleton (the stronger the smoothing, the shorter the filaments). We provide additional discussion on the filament length in  Appendix~\ref{app:subsampling}.
        
    \subsection{Coincidence}
        \label{sec:coincidence}
        In this section, we aim at 
        investigating the degree of the coincidence between the filamentary structures identified by using dark matter particles and galaxies. To quantify the overlap between the two skeletons, we use two different methods, one based on the "distance" between the skeletons of the two FSs, and another using additional information about critical points of the skeletons (maximas, minimas, saddle-points, bifurcation). In both cases, we use the DMFS, which we assume to be the ``real'' network,  as  reference  and quantify how much the GFS deviates from it.
        
    \subsubsection{Coincidence by distance between skeletons}
    \label{sec:coin_by_dist}
        The method we discuss in this section has been already used  in the literature \citep{Sarron+2019, Cornwell+2022}.  
        To quantify how much the two skeletons overlap we proceed as follows:
        for each segments of the reference DMFS skeleton, 
        we measure the distance to the nearest segment of the GFS skeleton\footnote{We obtain very similar results if we use the GFS as a reference, i.e. if we compute the distances of the DMFS from the GFS.}.
        If the structures were exactly the same, we would expect a distance between the skeletons close to zero for all the segments. In practice, also considering that the parameters adopted to determine filaments in the particles and galaxies distribution are not the same, we have to allow for some scatter, so we expect to obtain a unimodal distribution of  distances peaked close to zero. 
        \begin{figure}
            \centering
             \includegraphics[width=1\linewidth]{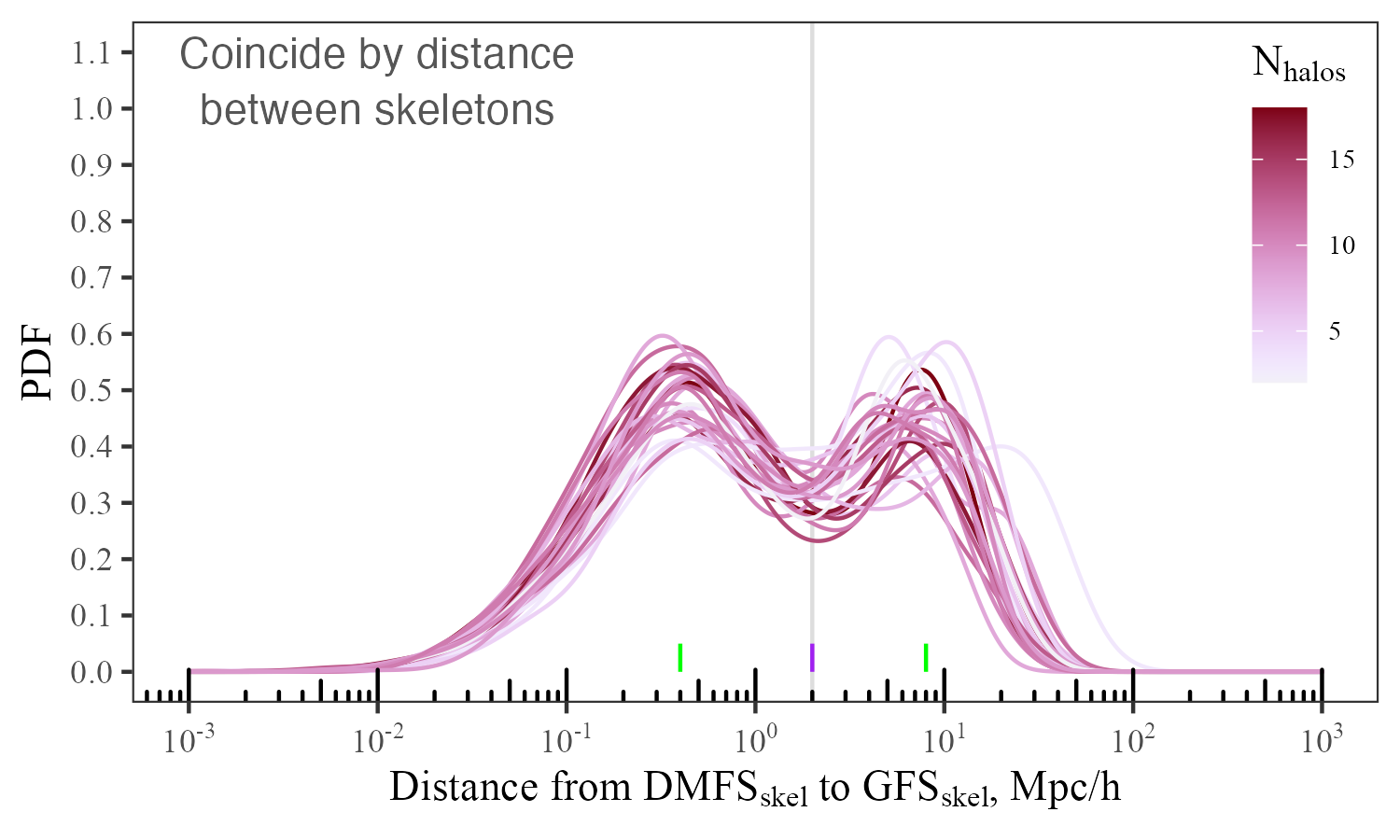}
              \includegraphics[width=1\linewidth]{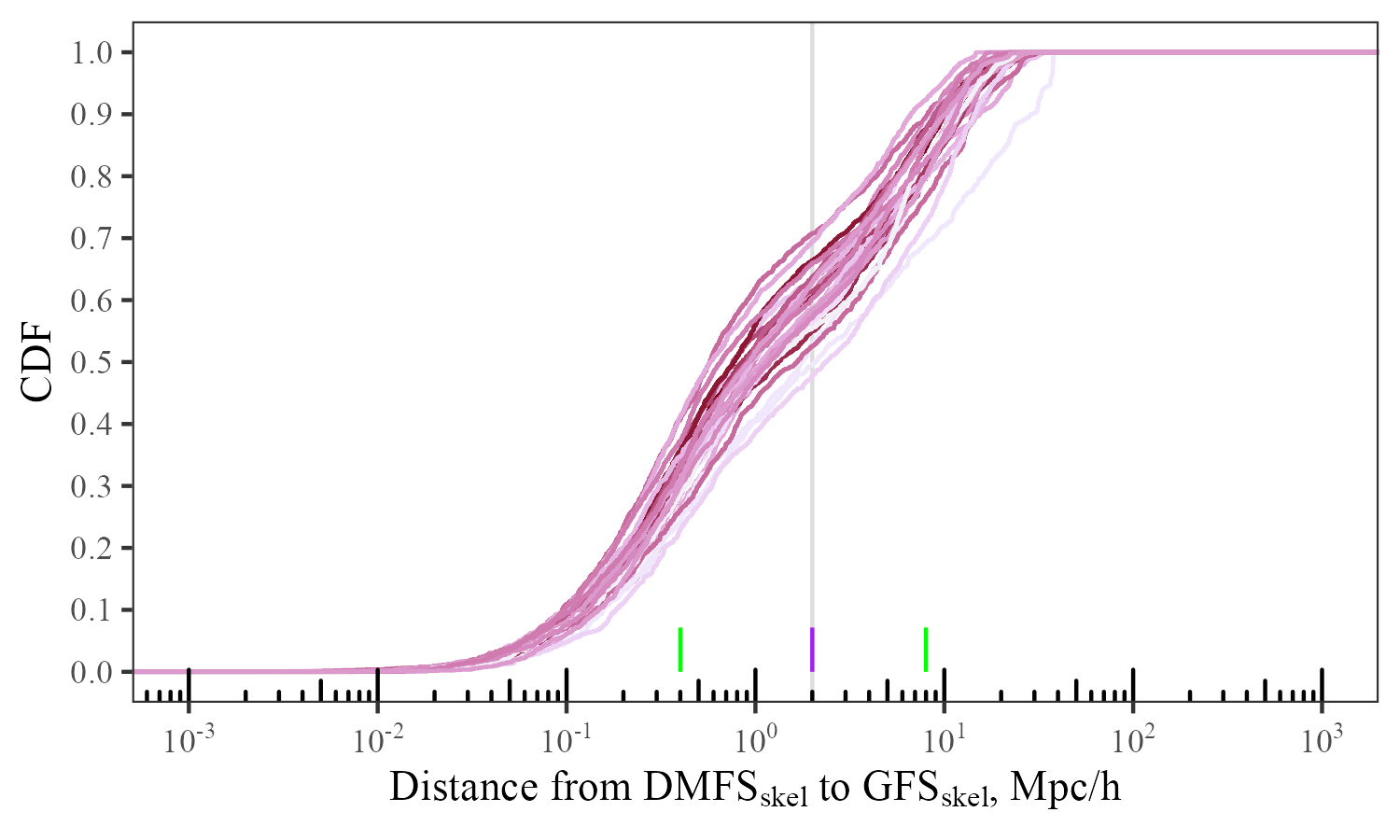}
            \caption{Coincidence by distance between skeletons: Top: probability distribution function (PDF) of the distance of all segments of the GFS from those of the DMFS. Each line shows the distribution obtained for one cube, with the intensity of the line color varying depending on the number of massive ($M_h > 10^{14} M_{\odot}$) haloes in the cube.
            The distribution shows two peaks at $d \sim 0.4$ and $8$~Mpc/h (marked with green ticks), and a minimum at 
            $d \sim 2$~Mpc/h (marked with a purple tick). Bottom: cumulative distribution function (CDF) of the same quantity. The gray vertical line separates the coincidence region from the non-coincidence region according to this method.}
    \label{fig:coin_by_dist}
        \end{figure}     
        The top panel of Fig.~\ref{fig:coin_by_dist} instead shows that, for each cube, the distribution of the distances of the GFS from the nearest DMFS is characterized by a bimodal distribution, with a first peak at about 0.4 Mpc/h, a minimum at about 2 Mpc/h and a second peak at 8 Mpc/h. The two peaks have similar height, and the distribution can be well approximated by the superposition of two  {log-normal} distributions. We assume that the first peak corresponds to the set of filaments that coincide in the DMFS and GFS, and the second one to the set of DMFS structures that do not have a corresponding GFS section. The similarity of the height of the peaks indicates that approximately only half of the structures match. More precisely, the percentage of the DMFS structure that has a  GFS segment closer than 2 Mpc/h (first mode) is 59$\pm$6\%. 
        \par
        The color intensity of the  lines in  Fig.~\ref{fig:coin_by_dist} reflects the number of massive haloes ($M_h > 10^{14} M_{\odot}$) in each cube. The figure shows that, in the cubes hosting a low number of massive haloes (light purple lines), the height of the left peak is significantly smaller than that of cubes hosting a larger number of massive haloes.  This behaviour is better seen in the bottom panel of Fig.~\ref{fig:coin_by_dist}, showing the cumulative distributions of the distances.
        Therefore, the DMFS and GFS have a stronger overlap in regions around massive haloes.
        To further support this claim, we check if the number of massive haloes in a cube has an impact on the results. Selecting  
         only the cubes with more than 12 massive haloes, we find that the  coincidence by distance between the GFS and DMFS skeletons improves (to 62$\pm$4\%), while selecting only the cubes with less than 6 massive haloes, it worsens (down 53$\pm$4\%). 
        Finally, we note that if we compute coincidence by distance between skeletons separately for the three sets of cubes (Coma-like, Virgo-like and randomly selected), we do not measure any statistical difference. This is due to the fact that in all these cubes a mix of haloes of different mass is present. 
       
    \subsubsection{Coincidence by coverage of DMFS critical points}
\begin{figure}
            \centering
             \includegraphics[width=1\linewidth]{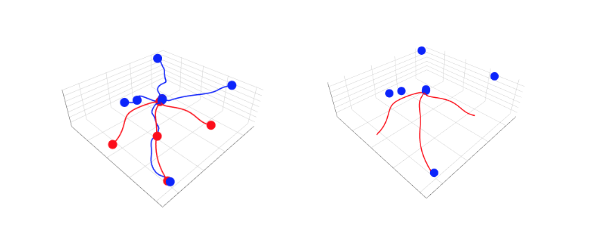}
            \caption{Schematic illustration of the coincidence by coverage of critical points method. The left image shows two distinct filamentary structures consisting of critical points and integral lines. The right image shows the red segments that intersect at least one critical point of the other structure. In this example, the  coincidence by coverage of blue points
            is 33\%.}
            \label{fig:coin_crits_example}
        \end{figure}

        The method presented above relies on the knowledge of the integral lines obtained by DisPerSE, ignoring the information about the nodes of the network. 
        In this section, we present an alternative  method for comparing the two skeletons, 
        taking into account both pieces of information. 
        We define coincidence as the fraction of critical points of the DMFS skeleton that are covered~(crossed) by the integral lines of the GFS\footnote{Similar results are obtained when inverting DMFS and GFS in the analysis.}. A simplified illustration of this method is shown in Fig.~\ref{fig:coin_crits_example}. The  image on the left shows two DisPerSE-like structures consisting of a set of critical points and segments. The  image on the right shows only the red segments that intersect at least one critical point.
        Only 2 out of 6 blue critical points are intersected by red segments. That is, the coincidence by coverage of blue points is 33\%. 
        
         \begin{figure}
            \centering
             \includegraphics[width=1\linewidth]{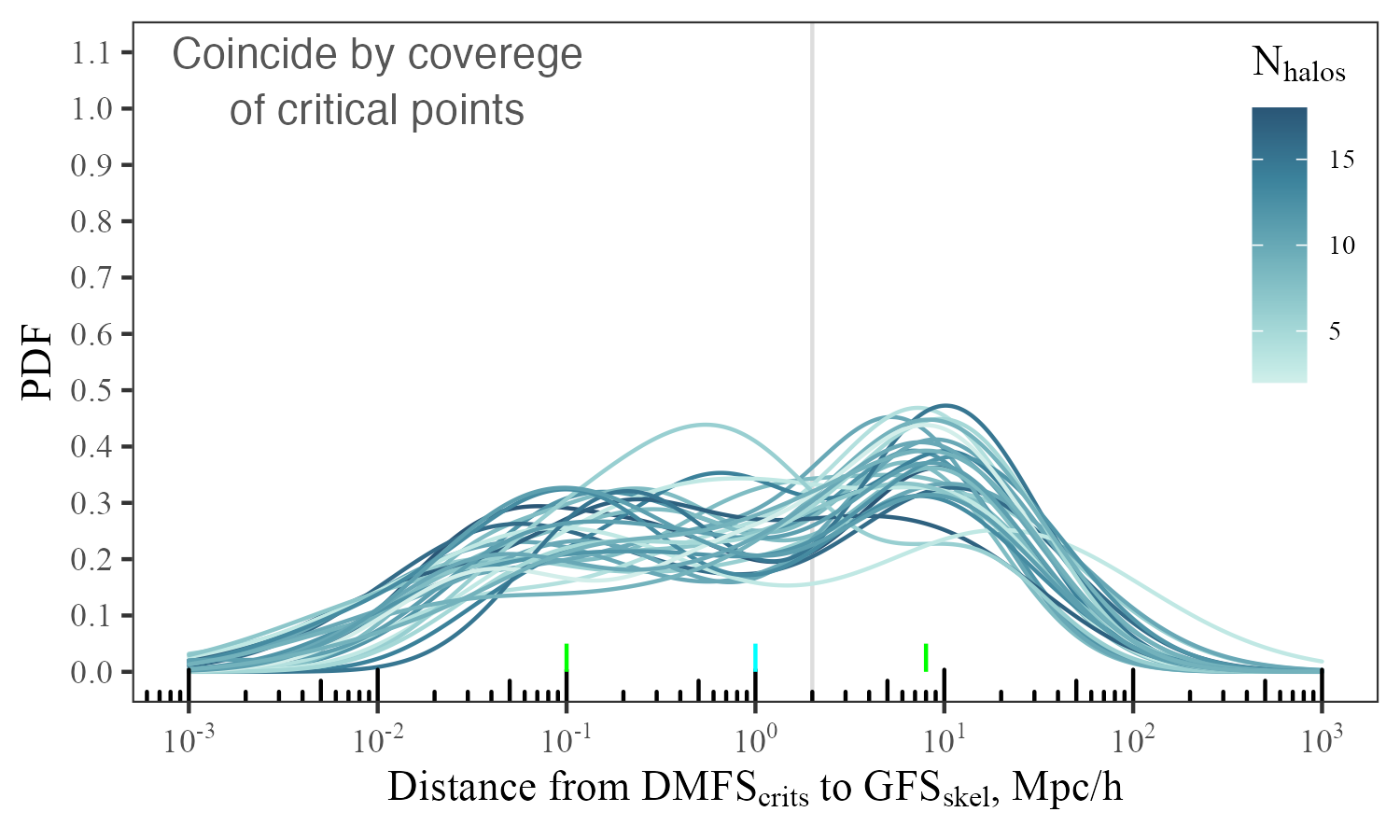}
              \includegraphics[width=1\linewidth]{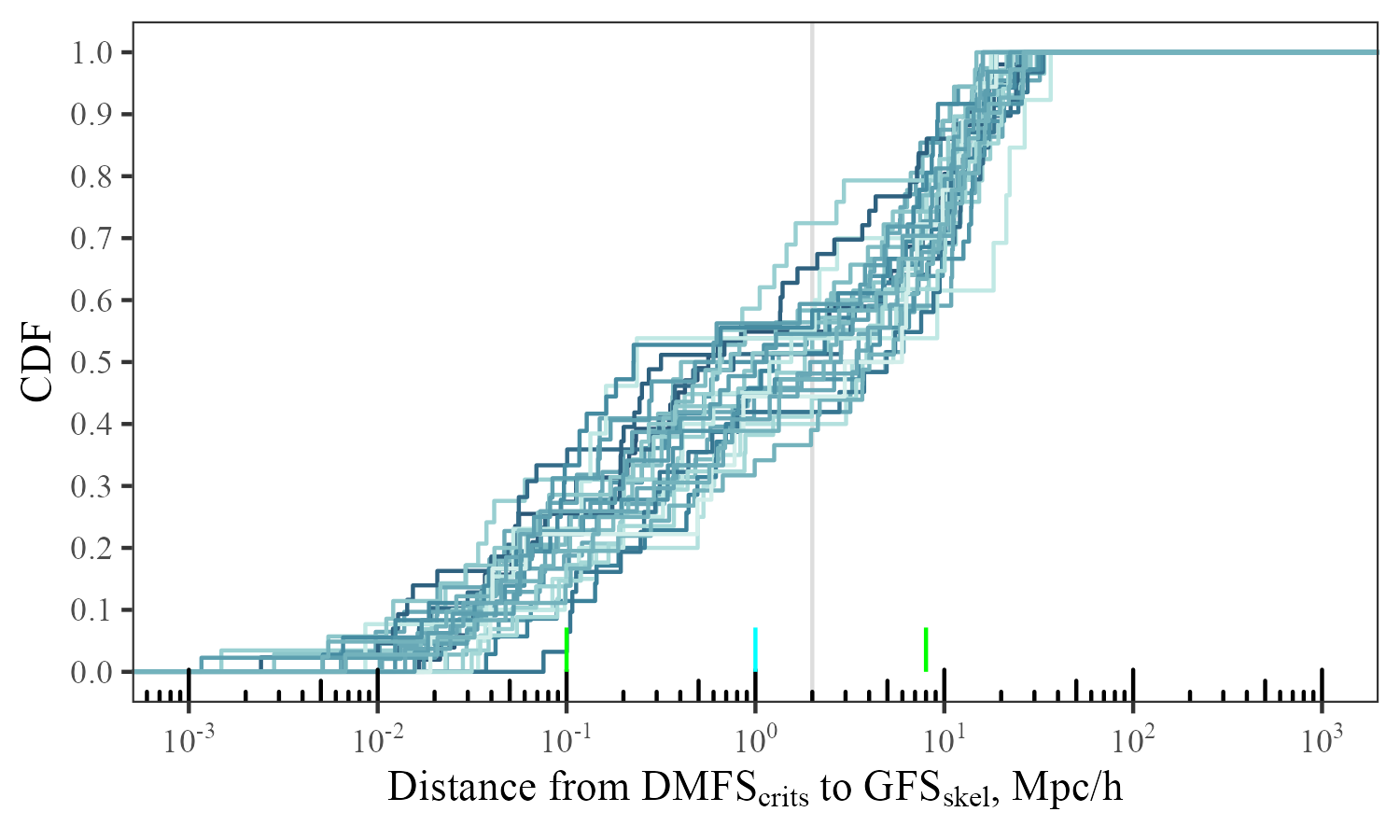}
            \caption{Coincidence by coverage of DMFS critical points. Top: probability distribution function (PDF) of the distances of each critical point  of DMFS skeleton to the nearest point of the GFS skeleton. Each  line represents one cube, with the color intensity of the lines varying depending on the number of massive haloes in each cube. Bottom panel: corresponding cumulative distribution functions (CDF). The tick segments show the location of the first and second peak positions, as well as their separation. The gray vertical line separates the coincidence region from the non-coincidence region according to this method.}
            \label{fig:coin_by_crits}
        \end{figure}
        
        \par
        As for the previous method, we  measure
        the distance between each critical point of the DMFS and the nearest segment of the GFS. The distribution of these distances is  shown in the top panel Fig.~\ref{fig:coin_by_crits}. Similarly to Fig.~\ref{fig:coin_by_dist}, the distribution  is  bimodal. 
        The first peak is located at less than 0.1 Mpc/h for the vast majority of the cubes considered, indicating  that the critical points of the DMFS are very close to the segments of the GFS (closer than the characteristic distance for  filaments identified by both skeletons, which is 0.4 Mpc/h according to the coincidence by distance between skeletons). The second peak is located at 9 Mpc/h. As for the previous method, we assume that this peak includes DMFS critical points that are not covered by GFS segments. 
        \par
        If we use the 2 Mpc/h division  to determine the fraction of critical points of one FS that are captured by the other skeleton, then Fig.~\ref{fig:coin_by_crits} shows that only about $\sim 52\pm8\%$ of all critical points of DMFS are matched by GFS segments in each cube.  
        \par 
        If instead of taking into account all the critical points we consider only those corresponding to the density peaks (maxima), we can estimate how the GFS captures the density peaks of the DMFS, rather than assessing the level of overlap between the two skeletons.  
        In this case, we find that the median  proportion of DMFS maxima that have a  GFS segment closer than 2 Mpc/h is $38\pm12\%$ for all cubes. This value is strongly dependent on the number of massive haloes in the cube: it raises to $45\pm14\%$ when considering only cubes with at least 12  haloes with  $M_{h} > 10^{14} M_{\odot}$, and drops to $33\pm3\%$ when considering only cubes with less than 6 such massive haloes. Results do not change when considering the GFS maximas captured by the integral DMFS lines: $38\pm11\%$ of galaxies density peaks are recovered. 

        To summarize,  we find that coincidence by distance between skeletons and coincidence by coverage of DMFS critical points provide consistent results, 
        highlighting that a non negligible fraction (about half) of the filaments obtained from the distribution of galaxies are not traced by overlapping dark matter filaments.

    \subsection{Connectivity}
        \label{sec:connectivity}
        \par
        Filaments can be seen as bridges connecting 
        haloes~\citep{Knebe+2004, Aragon-Calvo+2010}.
        Another way to quantify how well the different filament structures overlap 
        is to  count the  number of unique filaments that cross the virial radius of  haloes more massive than $10^{13.5} M_{\odot}$. This parameter is called connectivity and is a commonly used characterization of the cosmic web \citep{Codis+2018, Darragh_Ford+2019}.
        To compute the connectivity, we do not require that the filament necessarily begins inside  haloes, as we assume that a simple crossing of the virial radius can already impact the evolution of the halo.
        About 69 percent of haloes with $M_{h} > 10^{13.5} M_{\odot}$ in all cubes are crossed by at least one filament, regardless of filament type (DMFS or GFS). The fraction is up to 90\% for massive haloes~($M_{h} > 10^{14} M_{\odot}$). However, some  haloes can be intersected by several different filaments and this number can be different depending on the tracer considered. 
        
\begin{figure}
            \centering
              \includegraphics[width=1\linewidth]{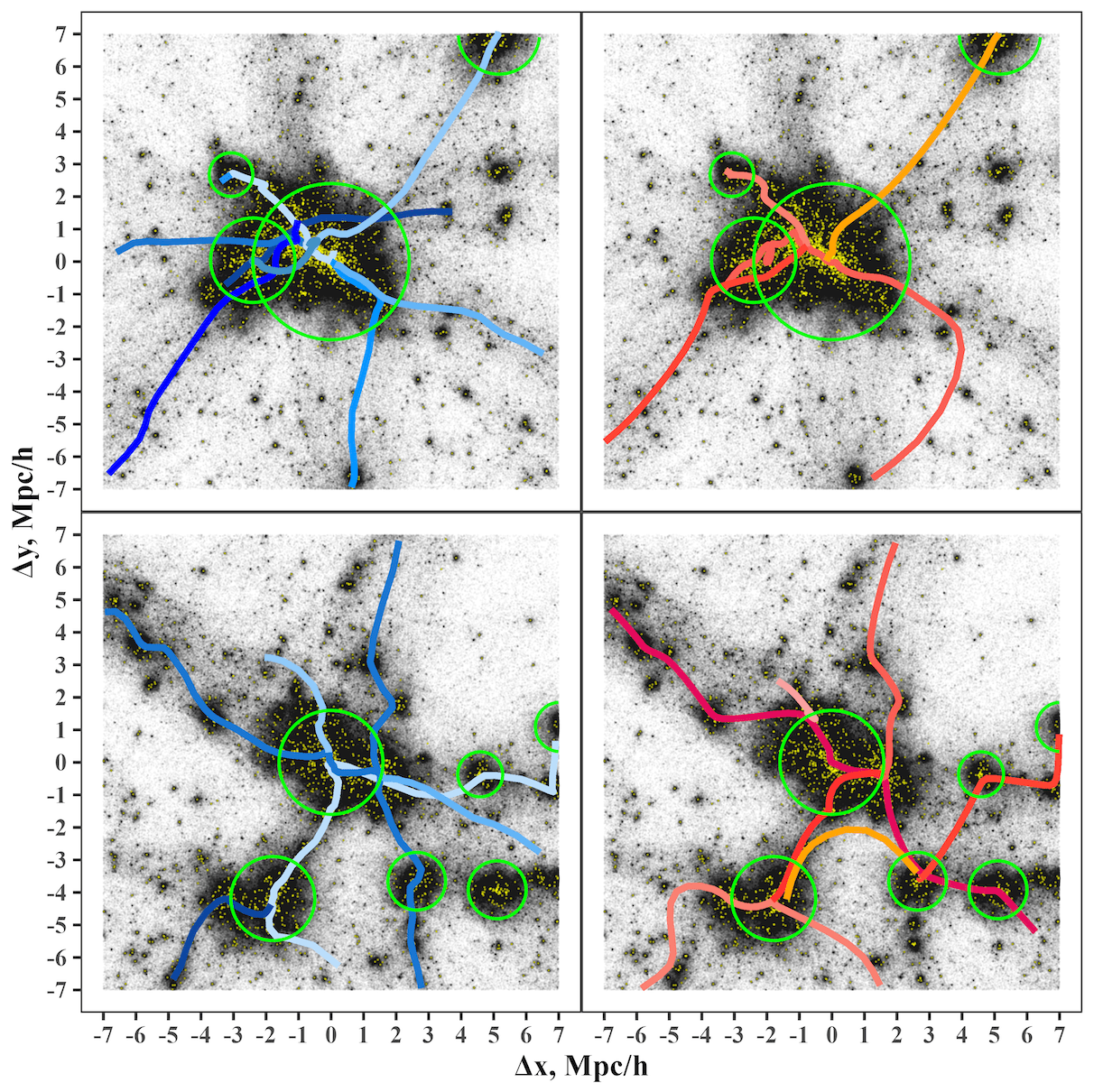}
            \caption{Slices of 7x7x7 Mpc/h around the most massive halo (Coma-like) in our sample with  mass $M_{h} = 1.6 \cdot 10^{15} M_{\odot}$~(top), and a Virgo-like halo with $M_{h} = 4.9 \cdot 10^{14} M_{\odot}$~(bottom). The black dots show the distribution of dark matter, while the yellow dots show the positions of galaxies. The green circles mark the virial radii of the haloes in the slices. The blue  {curves} in the left panel show the filaments identified when using dark matter particles as tracers, while the red  {curves} in the right panels are the filaments based on the galaxy distribution. A different shading is used for each filament.}
            \label{fig:connectivity_example}
        \end{figure}

        For example, in  Fig.~\ref{fig:connectivity_example} we show the regions around  two haloes with mass $M_{h} = 1.6 \cdot 10^{15} M_{\odot}$~(Coma-like, top) and $M_{h} = 4.9 \cdot 10^{14} M_{\odot}$ (Virgo-like, bottom), with highlighted the DMFS on the left and the GFS on the right. It is immediately clear that the network of filaments crossing the virial radius vary depending on the tracer used, with different number of filaments detected around different haloes. 
        \par
\begin{figure}
            \centering
             \includegraphics[width=1\linewidth]{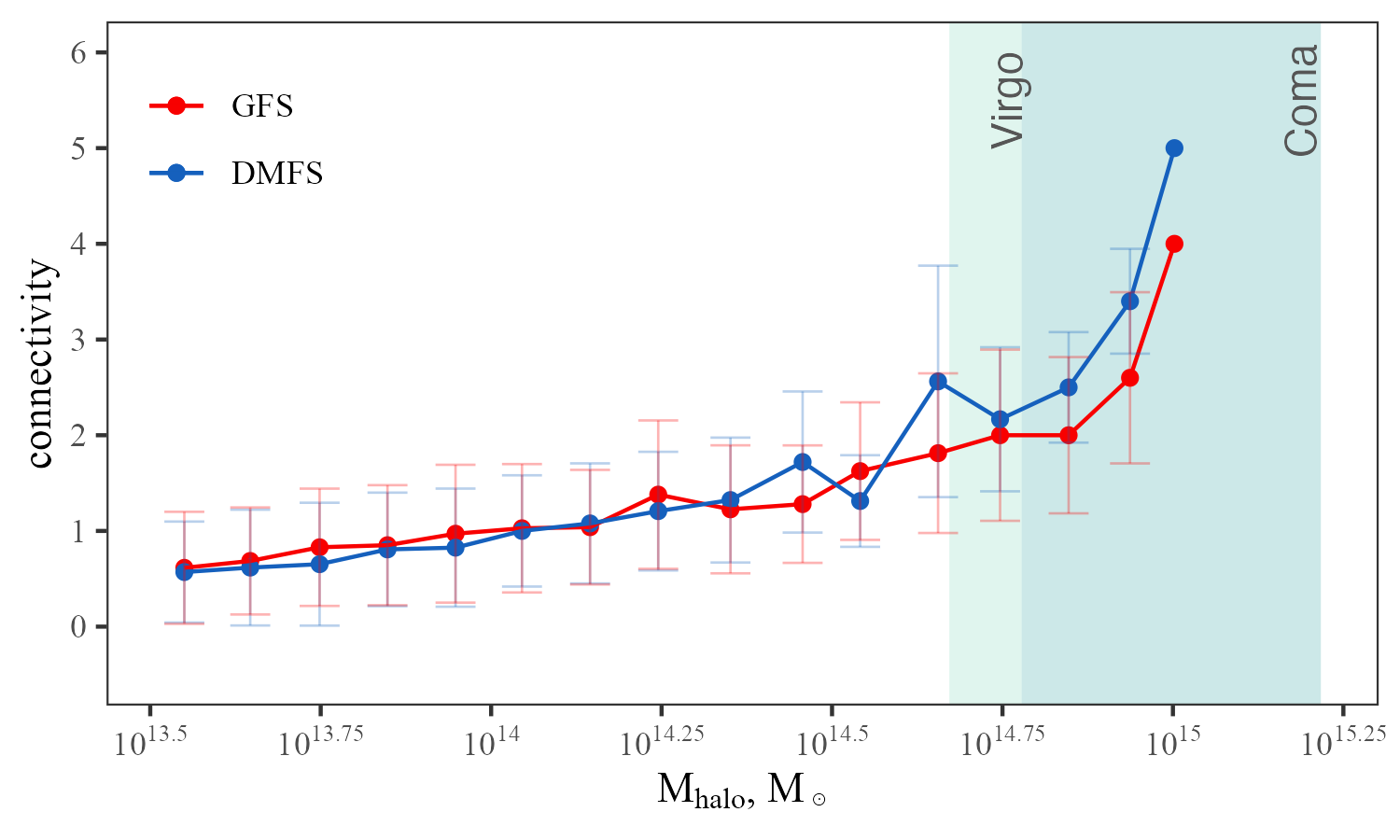}
            \caption{Distribution of the mean number of filaments crossing the virial radius  of a node  ($M_{h} > 10^{13.5} M_{\odot}$), 
             as a function of the mass of the halo. This  dependency is shown for DMFS~(blue) and for GFS~(red). Error bars show the standard deviation in each mass bin. The mass ranges of Virgo-like and Coma-like clusters are marked by the shaded regions.
            }
            \label{fig:connectivity_results_by_mass}
        \end{figure}

        The distribution of the number of individual GFS and DMFS filaments that cross the virial radius of haloes of $M_{h} > 10^{13.5} M_{\odot}$  is shown in  Fig.~\ref{fig:connectivity_results_by_mass}. The haloes of all cubes are parsed into mass bins and, for each bin, we calculate the mean and the variance.
        \par
        Regardless of the adopted tracer, the connectivity depends on the mass of the halo: the more massive the structure, the larger the number of filaments crossing its virial radius.  Haloes less massive than Virgo are typically associated with one or no filament. Virgo-like haloes have one to three connected filaments, depending on the type of tracer used, and Coma-like haloes have even larger number of crossing filaments. For haloes with mass $ > 3 \cdot 10^{14} M_{\odot}$, a larger number of dark matter filaments are detected than galaxy filaments. On average, the  GFS connectivity is systematically lower than the DMFS connectivity by $\sim1$.
        \par
        Similar results, confirming the increase of connectivity with increasing mass of the halo,  have been found by other studies \citep{Darragh_Ford+2019, Gouin+2021}, albeit the values we find tend to be lower than those published in earlier work. This may be ascribed to the adopted threshold for considering filaments and detailing of FS.
        The  connectivity we find for the Coma-like clusters is in agreement with \cite{Malavasi+2019}, who find a median value of the connectivity of 2.5. 
        \par
        
    \section{Dependence of the galaxy filament structure on the properties of galaxies used as tracers}
    \label{sec:different_tracers}
    
    In the previous sections, we have compared filaments extracted from the distribution of dark matter and from the distribution of galaxies with masses $M_{*} > 10^{9} M_{\odot}$.  
    Different choices could lead to significantly different results for what concerns the filament length, coincidence and connectivity. In what follows, we will inspect how the filament structures and some of the properties discussed above depend on the use of different tracers, and if/how much this affects the differences between GFS and DMFS that we have quantified in the previous section. Specifically, we use the GAEA semi-analytic model to test how results depend on galaxy stellar mass and what is the influence of `orphan' galaxies, i.e. those galaxies that are no longer associated with distinct dark matter substructures. In fact, the naive expectation is that, by using only galaxies that are either centrals of their own dark matter halo or associated with a distinct subhalo, we should obtain a better overlap with the filamentary structure that is traced by dark matter particles.

    \subsection{Varying the stellar mass of the galaxy tracers}
        \label{sec:different_tracers_mass}
        It is well known that galaxies represent biased tracers of dark matter density, and it is now well established that galaxy bias depends on galaxy properties, such as stellar mass~(for an overview, see e.g.,~\citealt{Desjacques+2018}). Recent work has also highlighted that massive galaxies tend to concentrate around filaments, while less massive ones are found both in dense regions and in the `field'~\citep{Kraljic+2018}. 
        In this section, we investigate how the GFS skeleton changes when taking into account only galaxies above a given stellar mass.
        In particular, we inspect whether the overlap between the DMFS and GFS improves 
        when considering galaxies with different stellar masses. 
        Specifically, we consider four different thresholds for the stellar mass ($M_{*} > 10^{7} M_{\odot}$, $M_{*} > 10^{8} M_{\odot}$, $M_{*} > 10^{9} M_{\odot}$~(GFS), and $M_{*} > 10^{10} M_{\odot}$)\footnote{Using galaxies $M_{\ast} < 10^{9} M_{\odot}$ is done for illustrative purposes, as the corresponding samples in GAEA are not fully resolved}.
    
            \begin{figure*}
                \centering
                \includegraphics[width=1\linewidth]{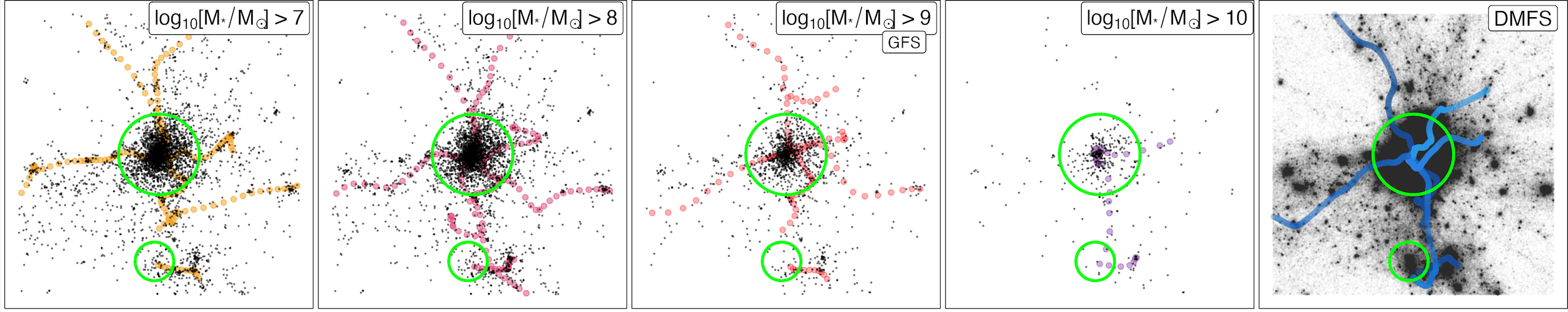}
                \includegraphics[width=1\linewidth]{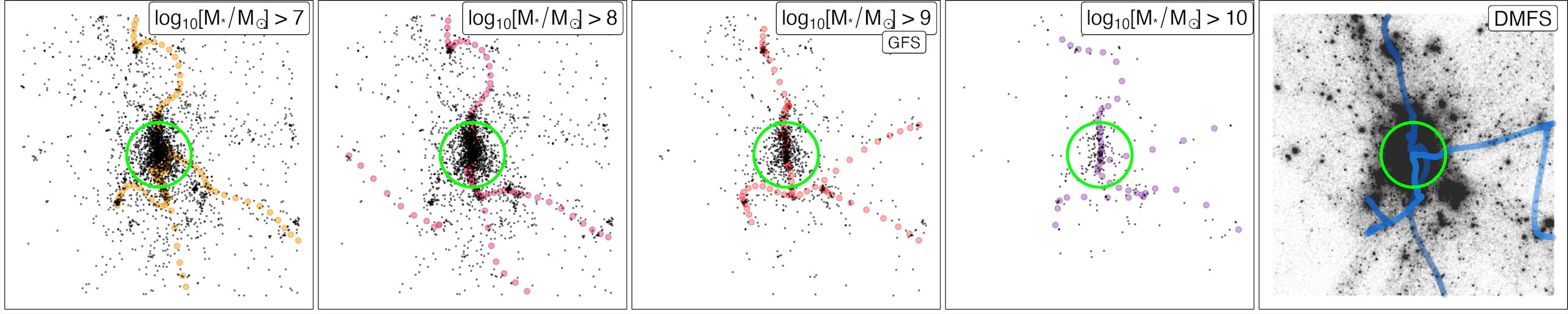}
                \includegraphics[width=1\linewidth]{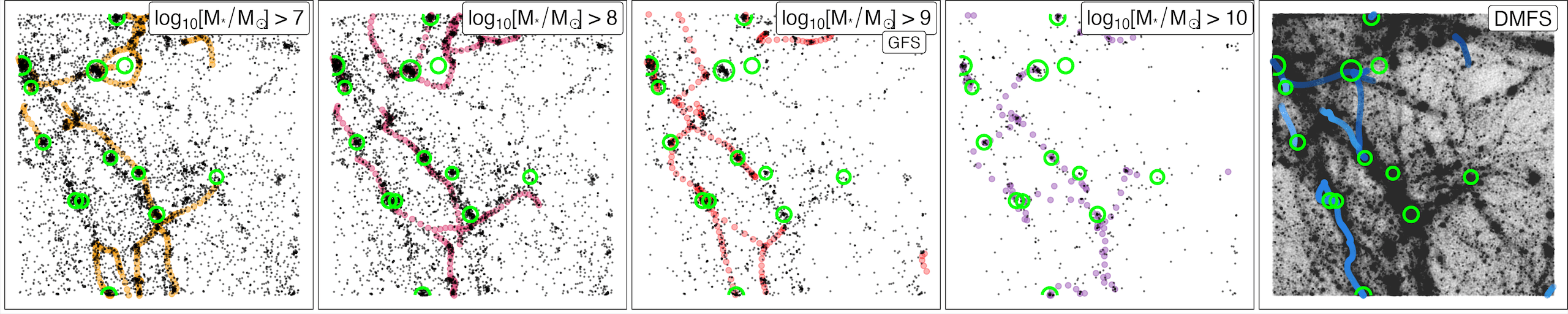}
                \caption{Filaments extracted applying different stellar mass thresholds.  The adopted stellar mass increases from left to right, as indicated in the labels. The rightmost panels show the dark matter distribution, for comparison. From top to bottom we show the extracted FS projected onto the XY plane around a Coma-like massive halo, around a Virgo-like massive halo, and in a random pointing, respectively. All regions shown correspond to slices of 14~Mpc/h on a side, but for the bottom region that instead corresponds to a slice of 30~Mpc/h.
                Black points indicate all galaxies above the adopted threshold, colored  {curves} the corresponding FS. ~ {The right column shows the distribution of DM particles with black dots.}
                Each green circle shows the virial radius of haloes more massive than $10^{13.5} M_{\odot}$.}
                \label{fig:fss_examples}
            \end{figure*}
        \par
        For each of the 27 cubes, and for each additional stellar mass cut considered, we run again DisPerSE and extract the corresponding filamentary structure, so that our condition on the total filament length is satisfied (the filament length - i.e., the sum of the lengths of all filaments - within each box should be approximately equal to that of the corresponding DMFS). Examples of regions around a Coma-like halo, a Virgo-like halo, and a region without any massive halo at the centre are shown in Fig.~\ref{fig:fss_examples}. 
        The figure shows that the FS extracted only using the most massive galaxies ($M_{*} >10^{10} M_{\odot}$) traces only the `strongest' filaments  in the dark matter distribution, for all three cases considered. 
        \par
        Lowering the mass threshold to $M_{*} > 10^{9}$ adds a significant number of galaxies, which also tend to aggregate around the densest dark matter regions.
        The corresponding GFS branches out significantly around massive clusters, getting much closer to the DMFS (see Sec.~\ref{sec:connectivity}, for a quantitative comparison). 
        \par
        Considering galaxies with  lower stellar mass ($ 10^{8} < M_{*} < 10^{9} M_{\odot}$), entails a change in the topology of the galaxy distribution within the cubes. Indeed, such galaxies are distributed both in filaments and in the  general field. As a consequence, we  both detect new filaments (trace  faint filaments that are not detected in the GFS using only more massive galaxies) and  observe a change in the detailed shape of the filaments when comparing them to those that are identified by using only more massive galaxies. 
        \par
        Going further down in stellar mass does not introduce any additional significant change. This is due to the small number (only $\sim 23\%$ of the galaxies with $M_{\ast} > 10^{7} M_{\odot}$) of galaxies with stellar mass $ 10^{7} M_{\odot} < M_{\ast} < 10^{8} M_{\odot} $ in the cubes, because this stellar mass is well within the resolution limit of our model applied to the Millennium Simulation.
        \par 
        This simple visual comparison shows that the stellar mass completeness of the observational sample play an important role. Including galaxies with stellar mass lower than $10^{9} M_{\odot}$ improves the overlap between the DM and galaxy filamentary structure identified, but only slightly. In the following, we quantify the visual comparison just discussed.
        
    \subsubsection{Coincidence by distance between skeletons}
\begin{figure}
            \centering
            \includegraphics[width=1\linewidth]{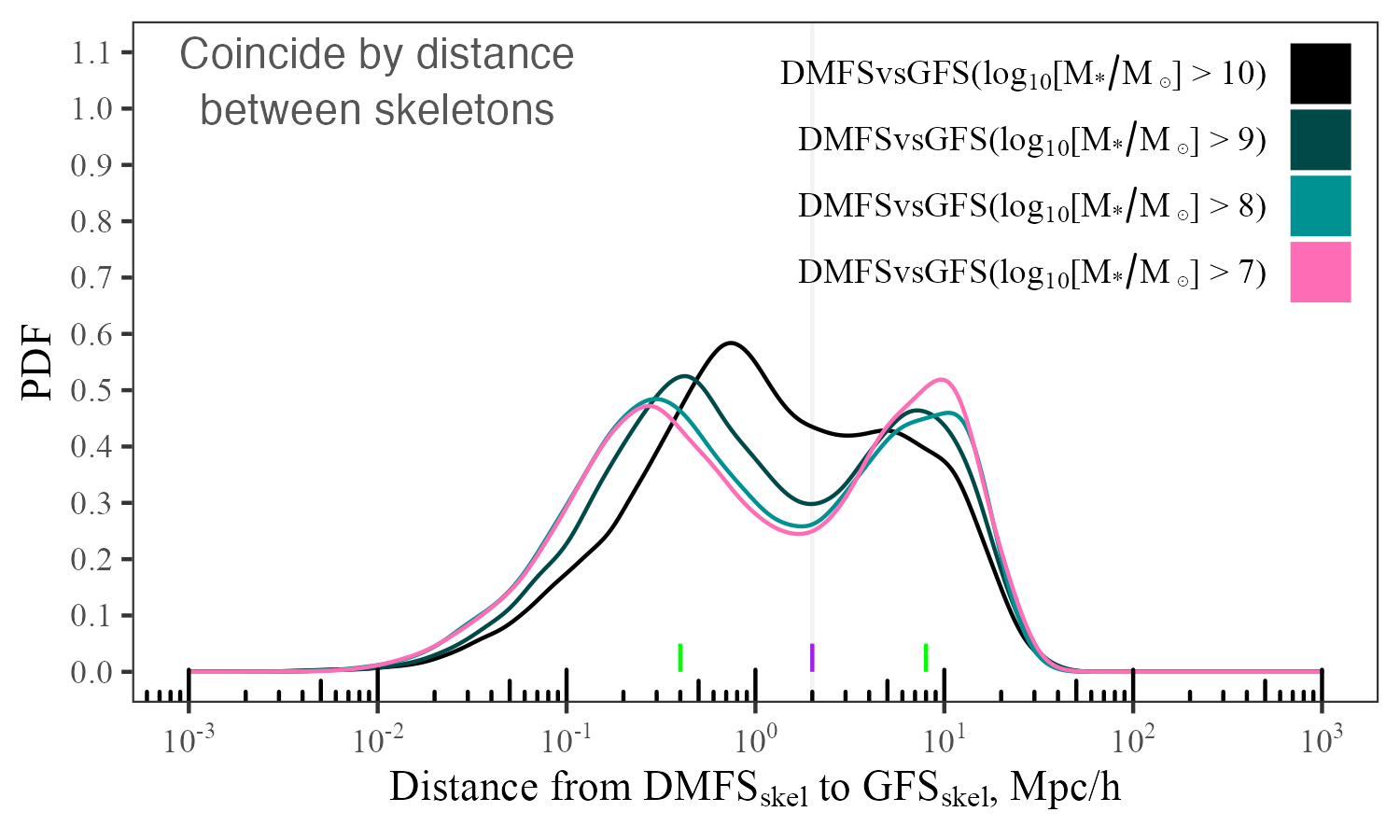}
            \caption{Coincidence by distance between skeletons of GFSs obtained adopting different  galaxy mass threshold and the DMFS. The tick segments
            were left from Fig.~3 and correspond to 0.4, 2, and 8 Mpc/h. Each line represents an average obtained considering all cubes.}
            \label{fig:diffmasscut_coin_by_dist}
        \end{figure}

        In order to support the conclusions based on the visual inspection of Fig.~\ref{fig:fss_examples}, we quantify how much the coincidence by distance between the DMFS  and the GFS varies when including galaxies of different mass. Fig.~\ref{fig:diffmasscut_coin_by_dist} shows the probability distribution function of the distance between the GFS obtained using different stellar mass cuts with respect to the DMFS; the reference case, corresponding to the GFS obtained using all galaxies more massive than $M_\ast = 10^{9} M_{\odot}$, is shown by the dark green line. Similarly to what was discussed in Section~3, the distances between the GFS and DMFS follow a bimodal distribution, for all samples considered.  The right peak corresponds to distances of about 9-10 Mpc/h, and represents the `non-coincident' parts of the skeletons. There is no significant improvement of the agreement between the GFS and the DMFS when including lower-mass galaxies. The fraction of DMFS segments that overlap with a GFS segment is always about 60 percent with differences around 1-3\% when considering different samples of tracers. It is noteworthy that the left peak of the distribution of distances obtained when including lower-mass galaxies  is shifted towards lower values,  indicating that lower-mass galaxies allow a slightly better tracing of the dark matter filamentary structure. This is connected to the fact that when low-mass galaxies are taken into account, the filaments contain a larger number of galaxies, which allows the galaxy density distribution to be defined with higher accuracy, and therefore a more accurate tracing of the filaments axes.
        \par
    \subsubsection{Connectivity}
        \label{sec:connectivity_diff_mass}
\begin{figure}
            \centering
            \includegraphics[width=1\linewidth]{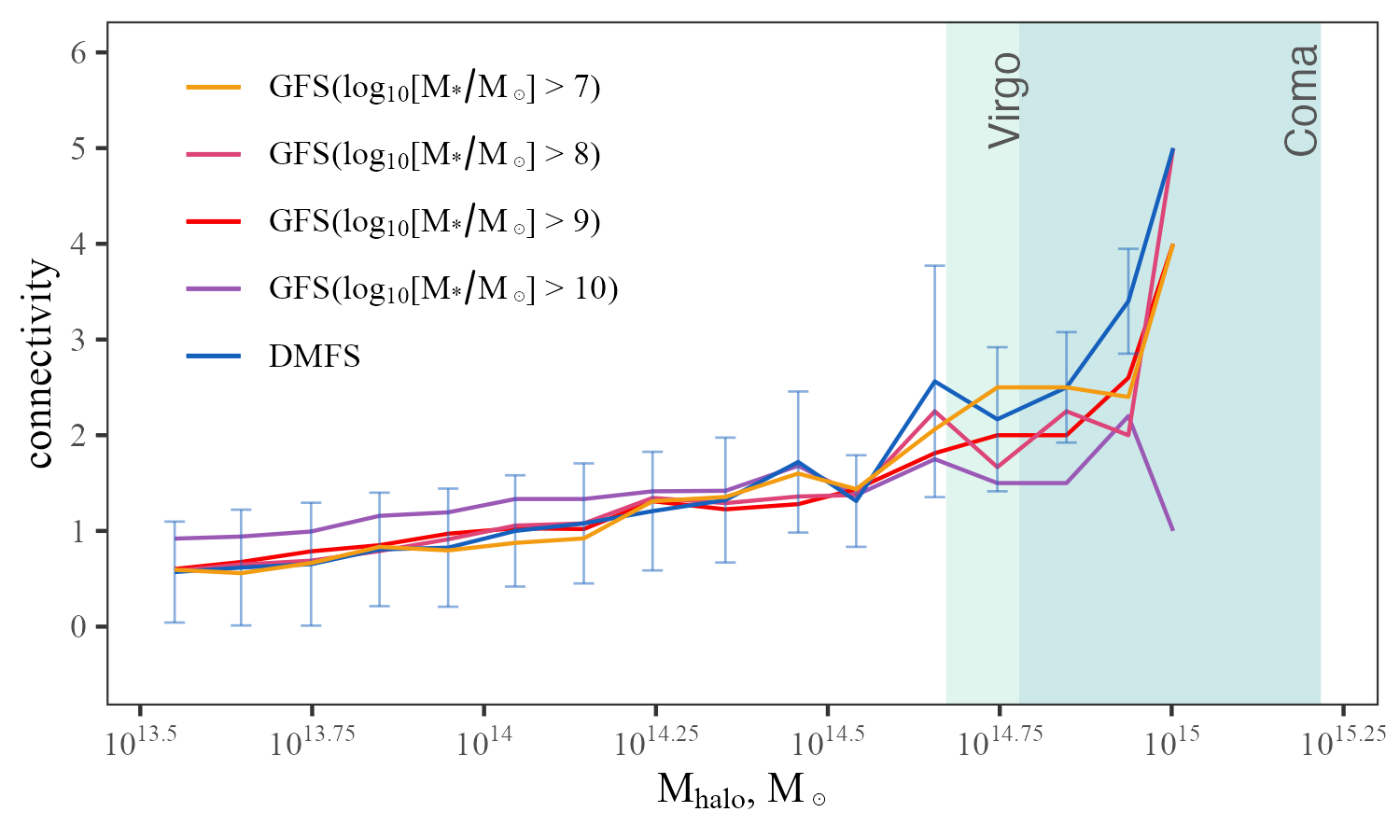}
            \caption{Connectivity as estimated considering different samples of galaxy tracers.  Errors bars are shown only for DMFS. For sake of clarity, we omit the other error bars, which have comparable  sizes ($\sim$2 for Virgo and Coma-like haloes, and $\sim$1 for lower halo masses).   }
            \label{fig:diffmasscut_connectivity}
       \end{figure}

        Next, we can also inspect how the inclusion of lower mass galaxies affects connectivity, as shown 
        in  Fig.~\ref{fig:diffmasscut_connectivity}. For haloes that are less massive than Virgo, we do not find significant differences between DMFS and all FSs that are extracted using galaxies less massive than  $10^{10} M_{\odot}$: one or no filament typically cross such haloes. The probability for a halo to be crossed by a filament increases with halo mass, and reaches 1 for haloes  with mass $3 \cdot 10^{14} M_{\odot}$~($10^{14.48} M_{\odot} $). The increase in connectivity for larger halo masses is not significant when considering the filamentary structure identified using only the most massive galaxies, confirming the visual impression of Fig.~\ref{fig:fss_examples}.
        Therefore, massive galaxies trace groups and/or clusters well, but not filaments. Evaluating the connectivity of massive haloes from the distribution of only massive galaxies would underestimate the result.
        \par
        Fig.~\ref{fig:diffmasscut_connectivity} shows that the inclusion of lower mass galaxies increases the values of the connectivity for all haloes more massive than $\sim 10^{14.4}\,{\rm M}_{\odot}$, bringing it closer to the connectivity that is estimated from the DMFS. We note, however, that this does not improve significantly the spatial coincidence of the filaments, as discussed in the previous section.
            
    \subsection{Centrals and satellites as tracers}
    
         Model galaxies in the GAEA simulation are classified into three types: centrals, satellites, and orphans. Centrals and satellites are associated with distinct dark matter subhaloes (centrals are associated with the `main halo', i.e., the bound part, of a friend-of-friend group), while orphan galaxies correspond to those whose parent subhalo has been stripped below the resolution limit of the simulation.  The model assumes that, being more concentrated than dark matter, galaxies can survive longer, and assign them a residual merging time that is parametrized using a variation of the Chandrasekkar dynamical friction formula \citep[e.g.][]{DeLucia+2010}. In GAEA, positions and velocities of orphan galaxies are traced by following the most bound particles of the subhaloes at the last time they could be identified. 
        \par
        As mentioned above, the naive expectation is that a better overlap with the DMFS can be obtained when considering only centrals and satellites, as they are associated with dense regions of the DM distribution. Moreover, it should be noted that orphan galaxies represent  $26\pm2\%$ of the total number of galaxies in our simulated cubes, so that uncertainties of position orphan galaxies can significantly affect identifying filaments.

\begin{figure}
            \centering
    
            \includegraphics[width=1\linewidth]{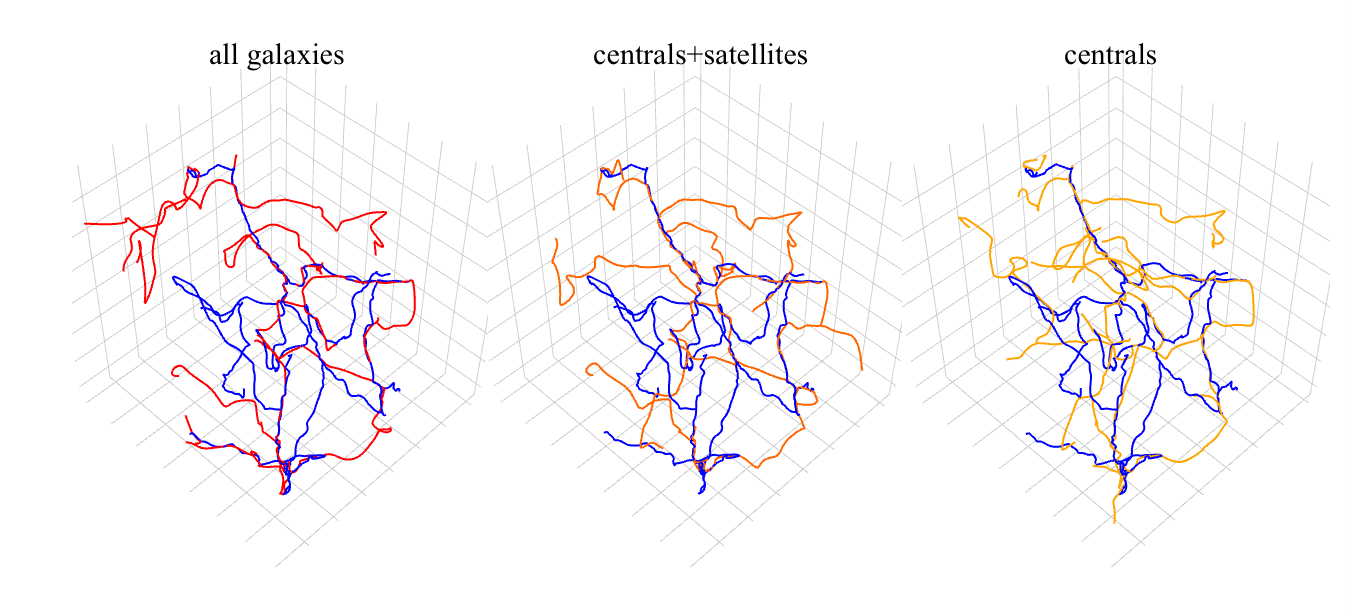}
            \caption{Example of the filamentary structure identified by all galaxies $M_{*} > 10^{9} M_{\odot}$ (left panel), centrals and satellites (middle panel), and only central galaxies (right panel). 
            DMFS are shown in blue color in each panel.
            The cube shown corresponds to 70 Mpc/h on a side.}
            \label{fig:fss_examples_types3d}
        \end{figure}

        To quantify the expectations just discussed, and the impact of orphan galaxies on the filamentary structure extracted, we consider two additional runs of DisPerSE obtained by using only centrals and satellites galaxies or only centrals (more massive than $M_{*}^{cen} > 10^{9} M_{{\odot}}$). As in all the other runs, DisPerSE parameters are set by imposing that the total filaments length is close to that obtained by considering the GFS based on the sample including all galaxies.  Fig.~\ref{fig:fss_examples_types3d} shows an example of the 3D and filamentary structure obtained when considering these different samples. 
    
    \subsubsection{Influence of `orphan' galaxies}
        \label{sec:orphan_gals}
\begin{figure}
            \centering
            \includegraphics[width=1\linewidth]{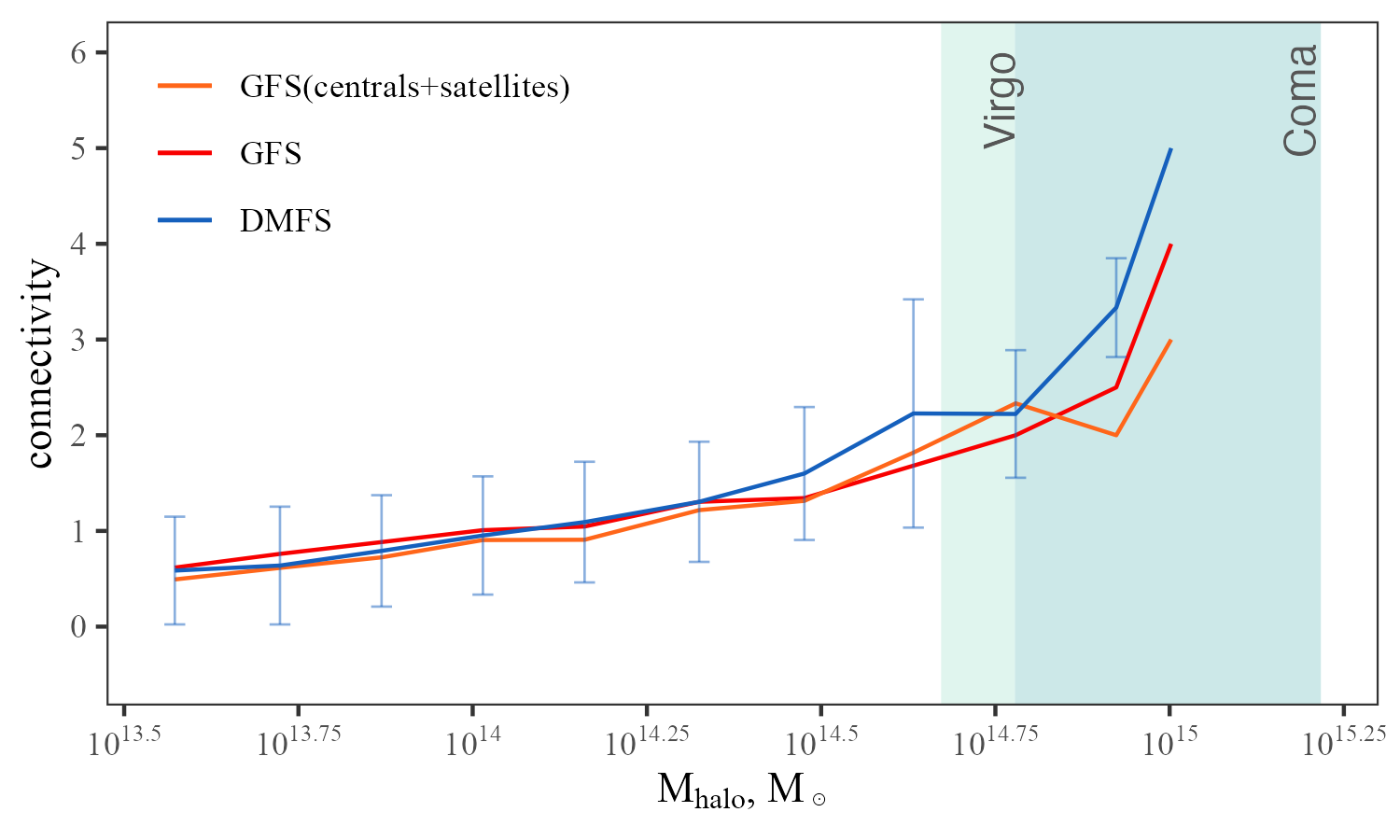}
            \caption{Connectivity as function of halo mass for GFS using all galaxies or only galaxies associated with dark matter haloes. As reference was added connectivity for DMFS and original GFS. Connectivity of FS only by centrals are not shown, because of centrals can not trace connectivity by construction. }
            \label{fig:fss_types_connect}
        \end{figure}

        Orphan galaxies tend to be found in large numbers in massive haloes, as can be appreciated from comparing the second and third columns of Fig.~\ref{fig:fss_examples_centrals}. 
        In general, we do not find significant differences in the filamentary structure identified when excluding orphan galaxies. This means that the central galaxies and their satellites (those associated with a distinct subhalo) identify filaments quite well: the coincidence value estimated in previous sections ($ 59\pm 6$ percent) reduces to $55\pm 8$ percent when we exclude orphans. In Fig.~\ref{fig:fss_types_coin} we quantify the overlap between the GFS and that of the dark matter when including/excluding orphans. The GFS obtained considering centrals+satellites trace the DMFS as well as the GFS obtained by including all model galaxies.
        The overall small impact of orphan galaxies on the FS identified is confirmed by the very small changes obtained for the connectivity value, as shown in Fig.~\ref{fig:fss_types_connect}. Only around the most massive haloes, where orphans become more important, one finds larger variations (still not very significant within the scatter) of the connectivity. 
        \par
        Given that the FS is not significantly modified when the orphan galaxies are removed, we can conclude that the central and satellites trace the distribution of dark matter as well as all galaxies. Additionally, the uncertainty on the positions of orphan galaxies does not affect extracting filaments definition even if orphans represent about one quarter of the total galaxy sample. We deem this to the fact that orphan galaxies are not randomly distributed, but over-abundant in dense regions like filaments and clusters.
        
    \subsubsection{Centrals as tracers} 
\begin{figure}
            \centering
            \includegraphics[width=1\linewidth]{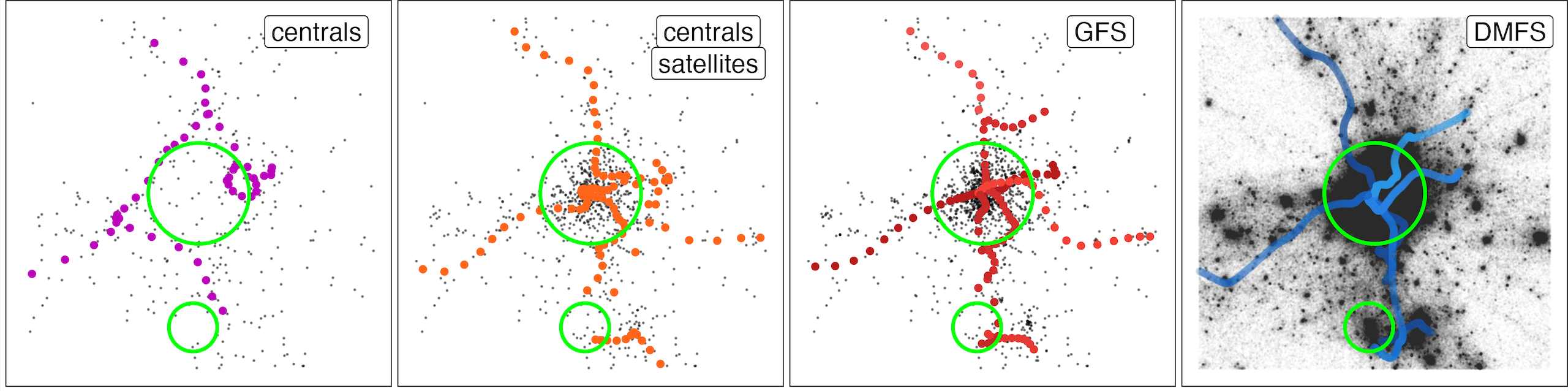}
            \includegraphics[width=1\linewidth]{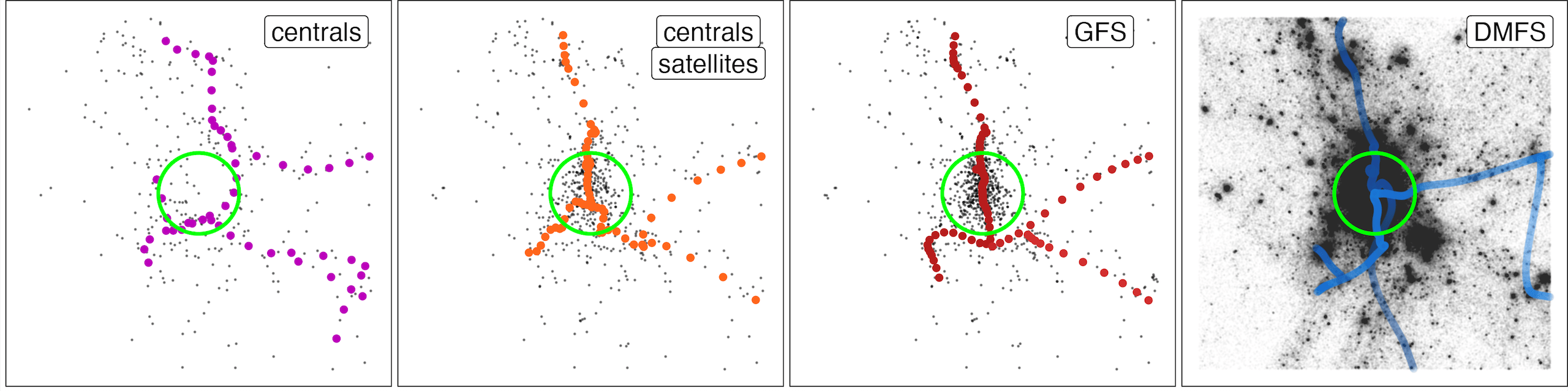}
            \includegraphics[width=1\linewidth]{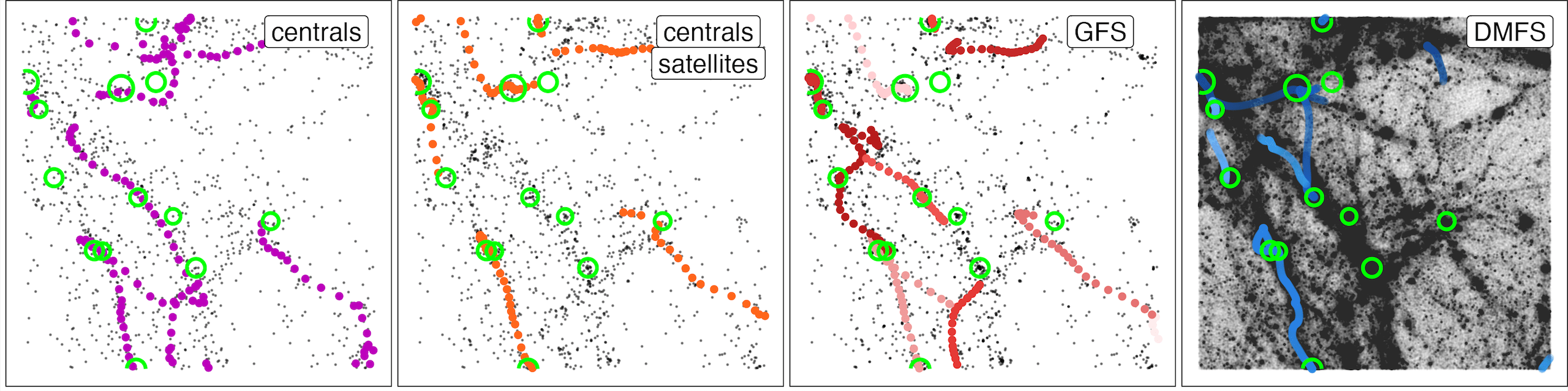}
            \caption{Filament systems extracted considering only central galaxies (first column), and centrals+satellites (second column). The fiducial GFS and DMFS are shown for comparison, in the third and fourth panels, respectively.  From top to bottom we show the extracted FS projected onto the XY plane around a Coma-like massive halo (the slice has a side of 14 Mpc/h), around a Virgo-like massive halo (the slice has a side of 7 Mpc/h) and in a random pointing (the slice has a side of 30 Mpc/h), respectively.
            Symbols and colors are as in Fig.~\ref{fig:fss_examples}.}
            \label{fig:fss_examples_centrals}
        \end{figure}

\begin{figure}
            \centering
            \includegraphics[width=1\linewidth]{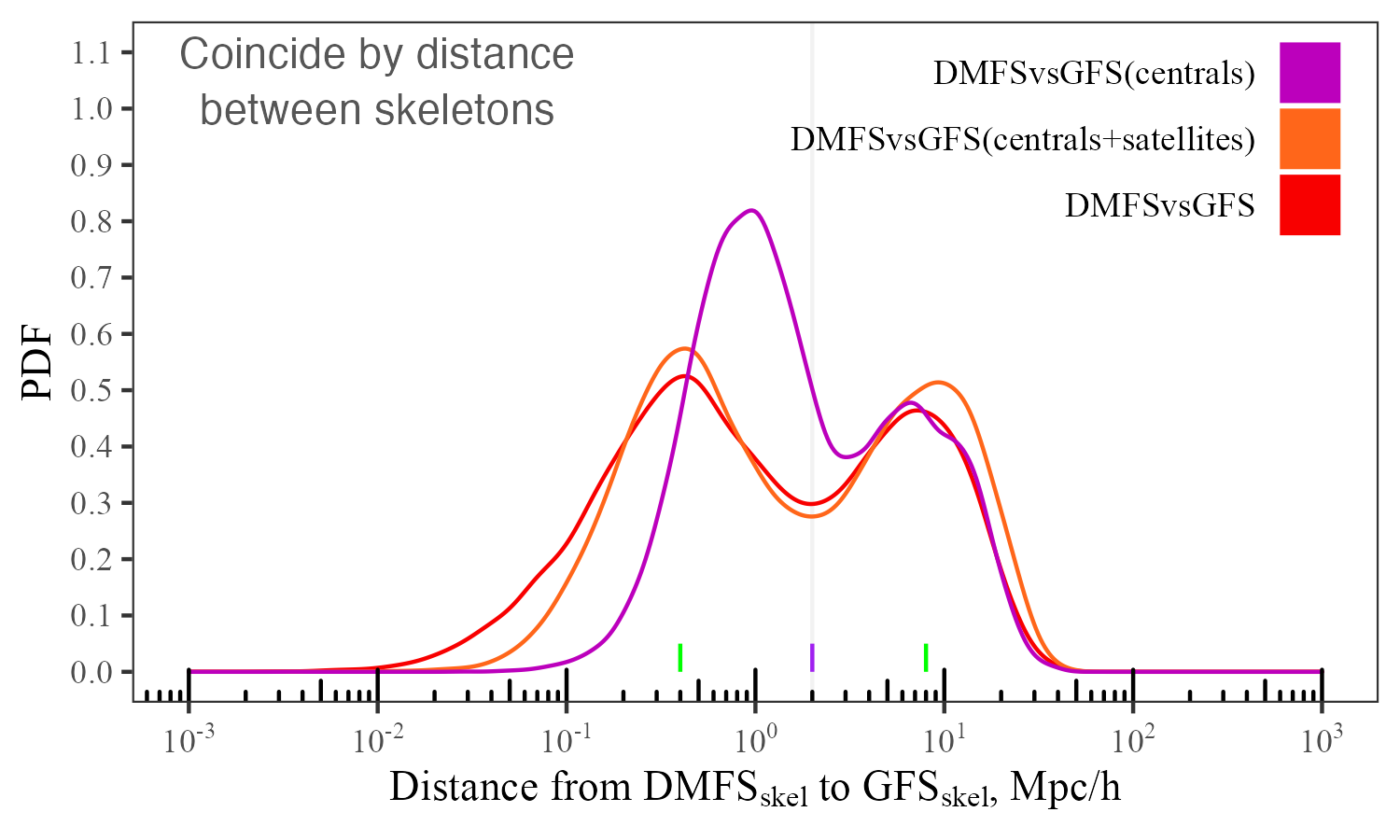}
            \caption{Coincidence by distance between DMFS skeletons and different GFS, which include all galaxies, only galaxies associated with dark matter haloes or only centrals galaxies. Serifs on the bottom according to the Fig.~\ref{fig:coin_by_dist}).}
            \label{fig:fss_types_coin}
        \end{figure}

        Figures~\ref{fig:fss_examples_centrals} and~\ref{fig:fss_examples_types3d} (first and third columns) show that centrals can trace the filamentary structures outside the virial radius of massive haloes, but can not trace (by construction) how filaments extend within the virial radius of haloes. Therefore, they cannot be used to define the connectivity. We note that dark matter filaments contain enough central galaxies inside to be distinguished as filaments in 70x70x70 Mpc/h$^{-3}$ volumes. The formal metric coincidence by distance is shown in  Fig.~\ref{fig:fss_types_coin}. Around $58\pm7\%$ of the GFS (centrals) segments can be found near~(close that 2 Mpc/h) the segments of the DMFS.
        Despite the fact that the absolute value of the coincidence has not changed significantly, we note a worsening of the tracing of the DMFS axis: the peak of the distance distribution is shifted from 0.4 Mpc/h to 1 Mpc/h. 

    \section{Summary and Discussion}
    \label{sec:discussion}

        In this work, we have applied the DisPerSE code to the outputs of the GAEA semi-analytic model to investigate 
        how the filaments extracted from the distribution of galaxies match  the filament structure obtained using dark matter particles. This analysis is  relevant for interpreting observational studies that do not have access to the distribution of dark matter. 
        To investigate how a biased galaxy distribution affects the identification of filaments at $z=0$, we used 27 simulated volumes: 9 are centred around Virgo-like haloes, 9 around  Coma-like haloes, and 9 corresponds to regions of `average' density, that have been randomly selected from a much larger simulated volume. For each of these cubes, we extracted the FS separately for the dark matter distribution and galaxies. We fine tuned the parameters of DisPerSE to obtain the same total length of the FS and cleaned the final structures excluding the 10\%  shortest filaments.  We introduced several metrics to quantify the overlap between the two structures:  filament length, coincidence by distance between skeletons, coincidence by coverage of the DMFS critical points, and connectivity. 
        \par
        Overall, DM filament are systematically longer than galaxy filament. The analysis of the coincidence showed that around 60\% of dark matter filaments have a corresponding filament in the structure identified by the distribution of galaxies with stellar mass $M_{\ast} > 10^{9} M_{\odot}$.  This result depends on the number of massive ($M_{h} > 10^{14} M_{\odot}$) haloes in the volume considered: for volumes containing less than 5 massive haloes, the overlap between the FSs traced by dark matter particles and galaxies gets worse. We also have shown that there is no significant effect on coincidence when uncertainties of position about a quarter of galaxy samples~(orphan galaxies) or using only centrals galaxies. To further extent, the coincidence between DMFS and GFS can not be improved by including low-mass galaxies ($M_{\ast} < 10^{9} M_{\odot}$) or using only massive ($M_{\ast} > 10^{10} M_{\odot}$) galaxies.  
        \par
\begin{figure}
            \centering
            \includegraphics[width=1\linewidth]{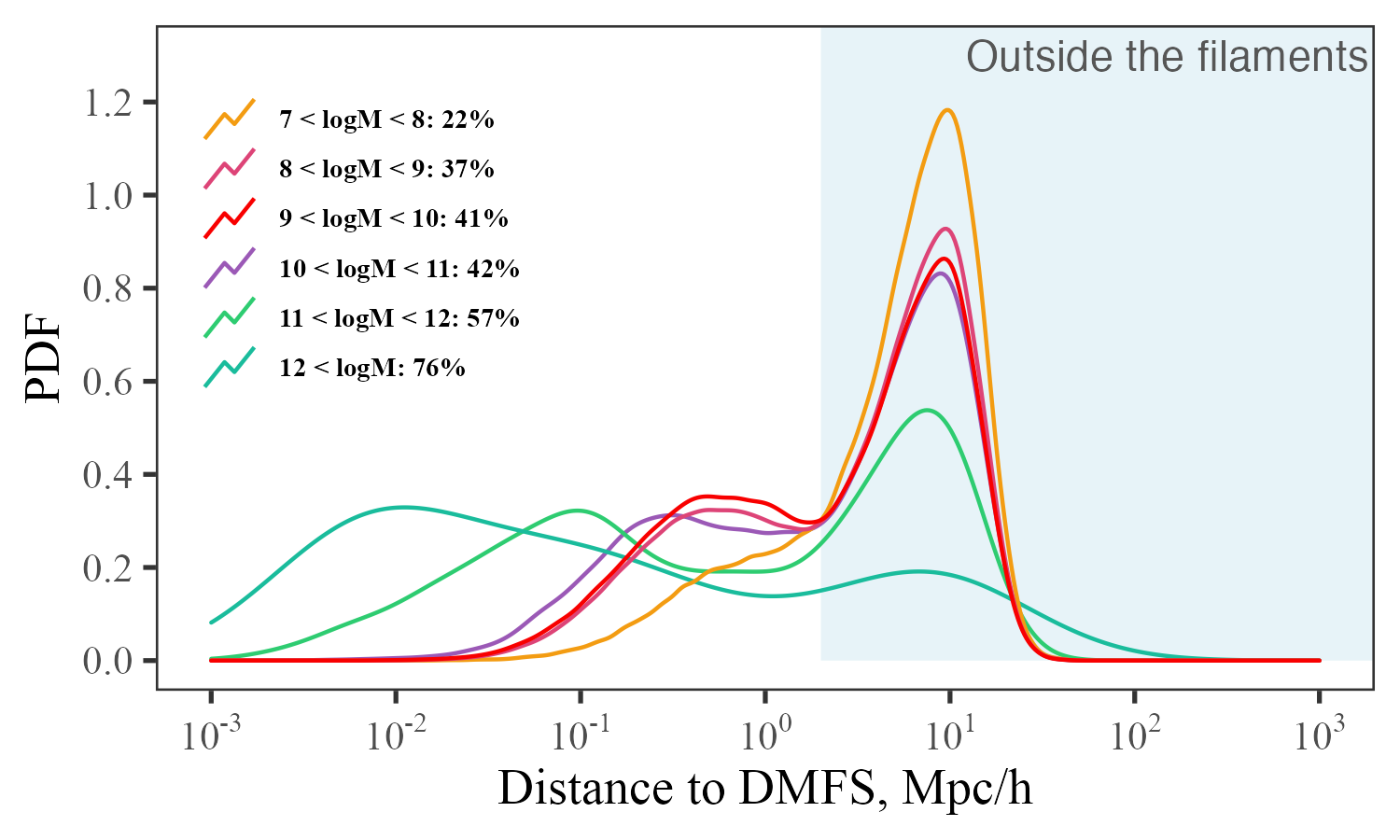}
            \caption{Probability distribution function of the distance to the DMFS axis for galaxies of different  mass, as indicated in the labels. 
            The shaded colored region indicates distances from the filament axis larger than 2 Mpc/h, which represents the radius of filaments, as obtained from Fig.~\ref{fig:coin_by_dist}. The legend also shows the fraction of galaxies of a given mass that are closer than 2 Mpc/h to the filaments.}
            \label{fig:dist_to_dmfs}
        \end{figure}

        To further support this result, we measured the distance from galaxies of different mass to the DMFS (Fig.~\ref{fig:dist_to_dmfs}). More than 70 percent of galaxies with $M_\ast>10^{11} M_{\odot}$ are at a distance of less than 2 Mpc/h from DMFS~(i.e. they belong to the filaments). In contrary, only 22 percent of $10^{7} > M_{\ast} > 10^{8} M_{\odot}$~( and 36\% of $10^{8} > M_{\ast} > 10^{9} M_{\odot}$) galaxies are that close to  filaments. Hence, it appears that massive galaxies not only are preferentially found in clusters and groups, but also in filaments, at least  in the local Universe, in agreement with previous results \citep{Laigle+2017,Kraljic+2018,Sarron+2019, Einasto+2022}. In contrast, low-mass galaxies are located not only near the filament, but also in regions with lower average density, introducing noise in  the filament extraction. These results should be interpreted with caution given that the galaxy catalogues we use in our study becomes incomplete for galaxy masses lower than $M_{\ast} \sim 10^{9} M_{\odot}$. 
        \par
        Considering connectivity, we found that overall 65\% of all haloes more massive than $10^{13.5} M_\odot$ are crossed by at least one filament, regardless of the adopted tracer and of the galaxy stellar mass cut adopted to extract filaments. The connectivity though depends on the mass of the halo: the more massive the structure, the larger the number of filaments crossing its virial radius.  
        \par
\begin{figure}
            \centering
            \includegraphics[width=1\linewidth]{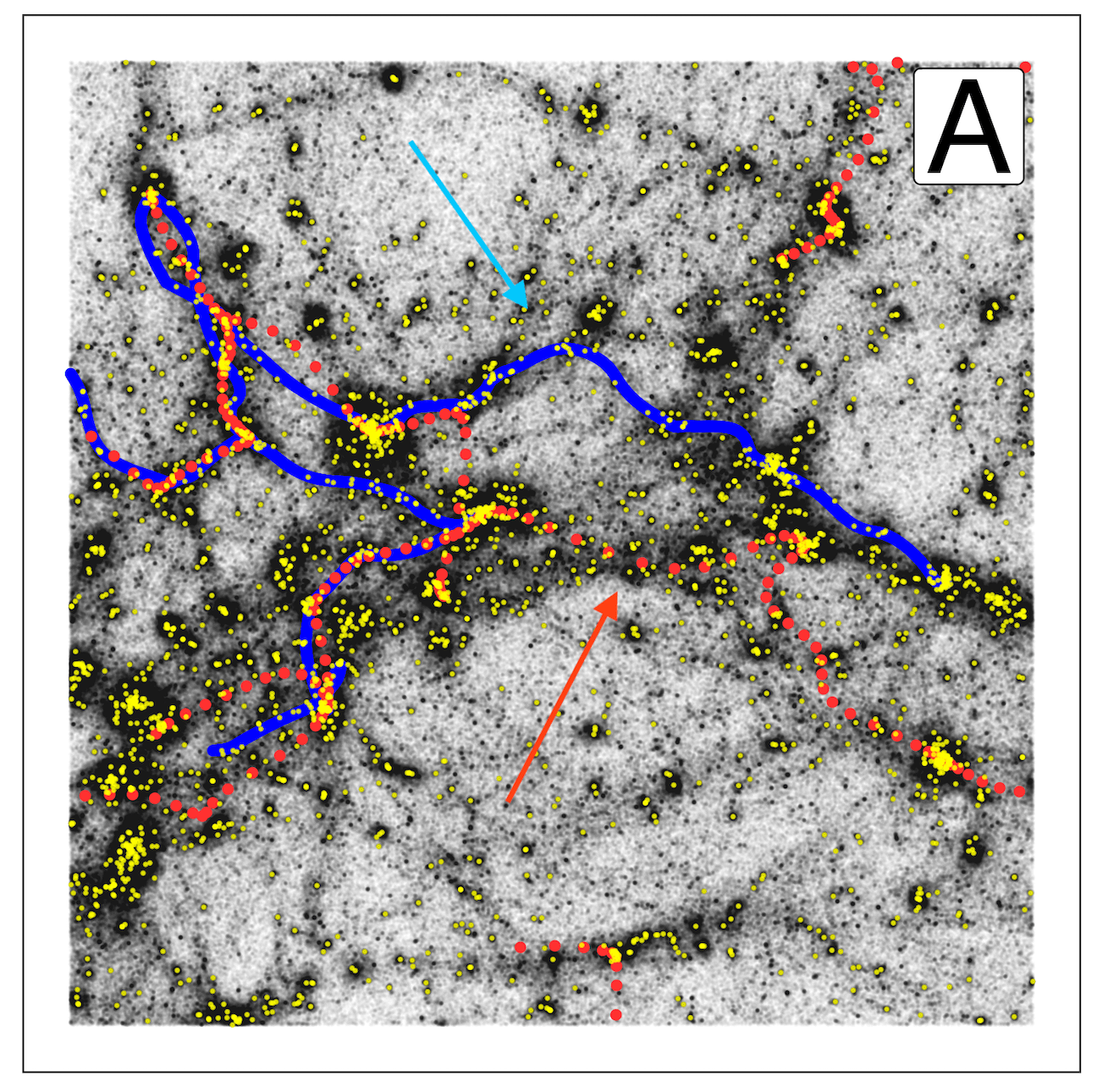}
            \includegraphics[width=0.48\linewidth]{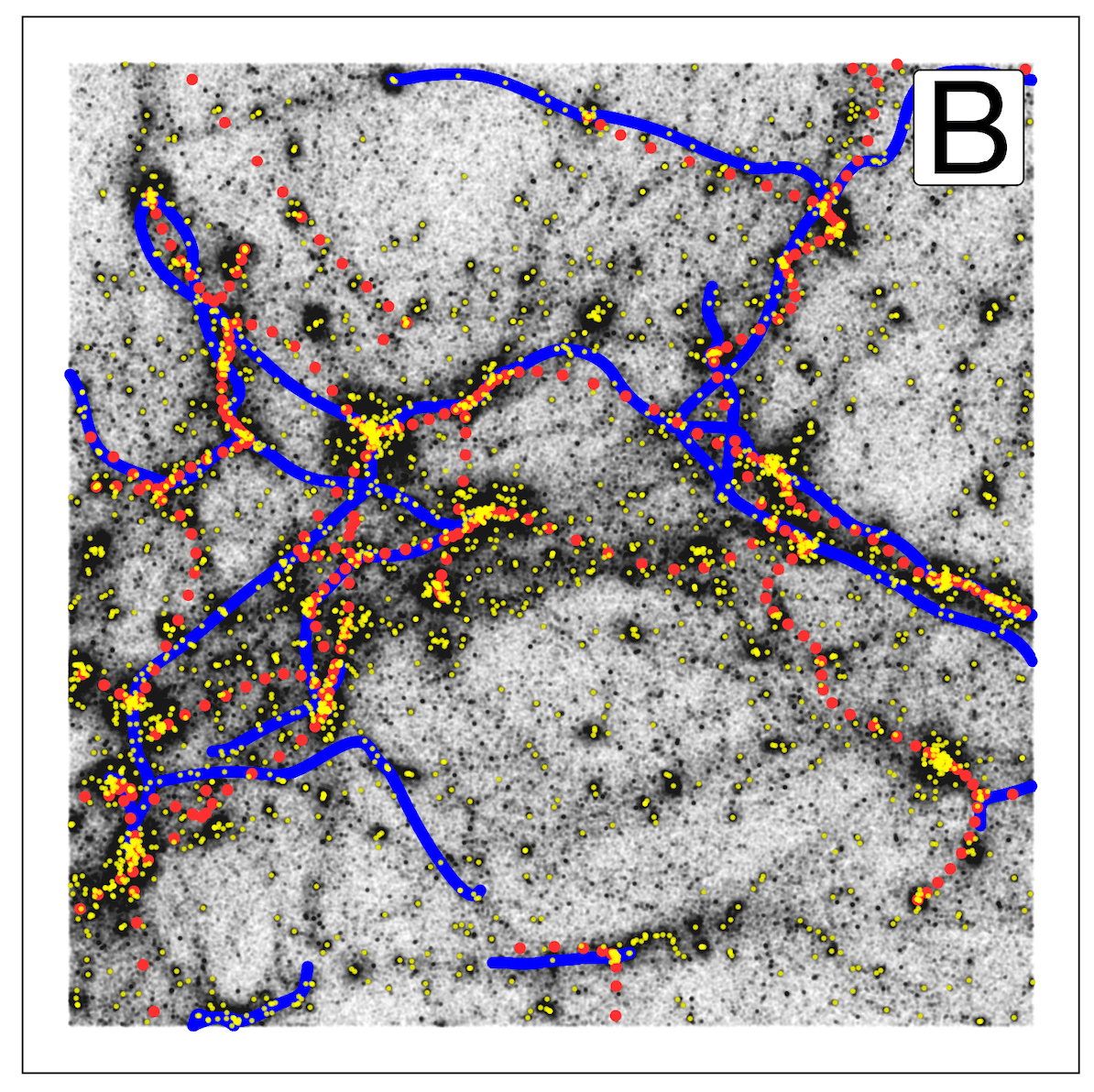}
            \includegraphics[width=0.48\linewidth]{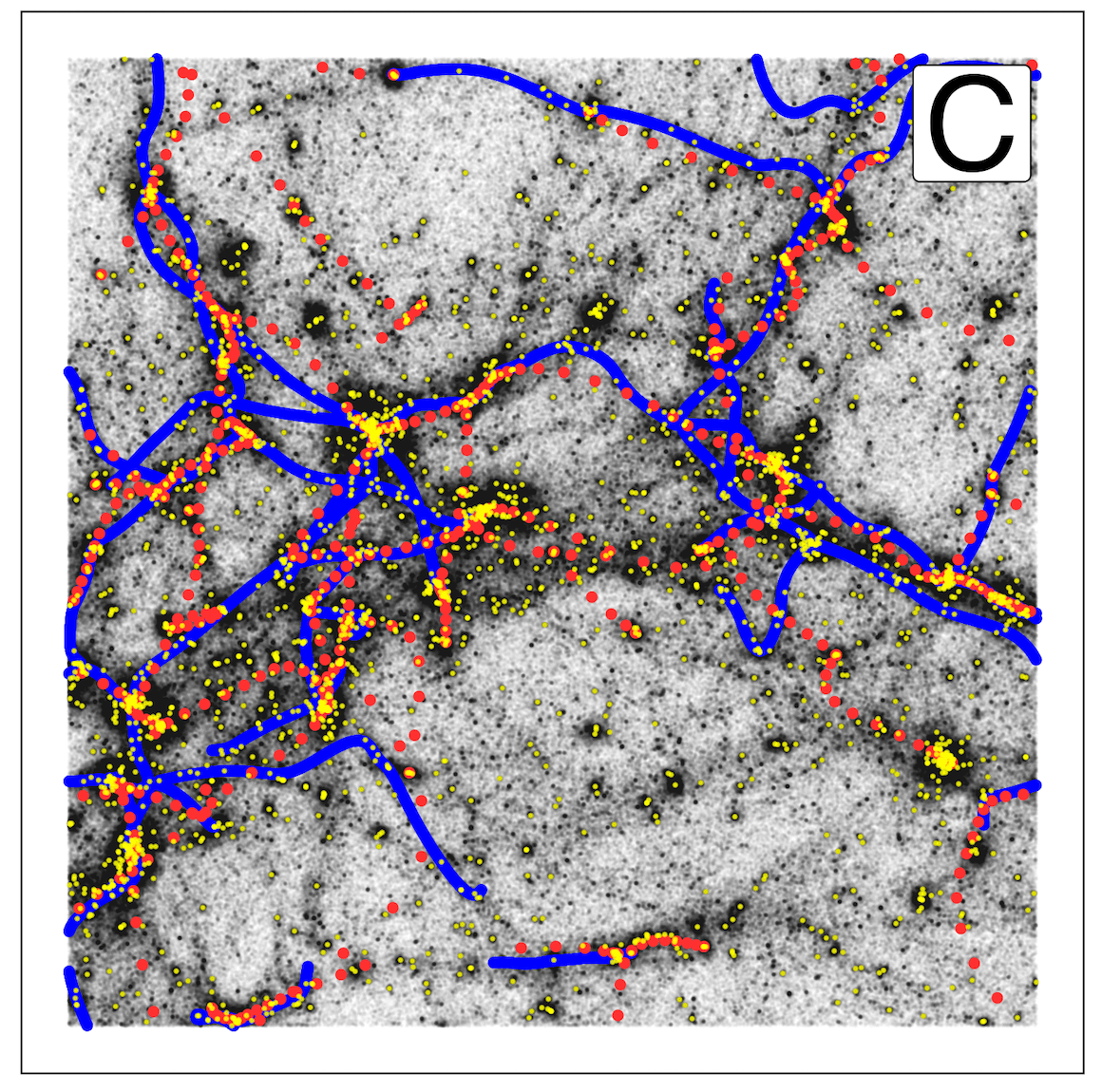}
            \caption{Panel A shows the DMFS (blue) and GFS (red) extracted using the thresholds adopted in main part of work~( {$10^{9}$} and  {$10^{4}$}, respectively).  Panel B (C) shows the results corresponding to lower threshold values for DMFS --  {$5 \cdot 10^{8}$} ( {$3 \cdot 10^{8}$}) and for GFS --  {$10^{3}$} ( {$10^{1}$}). Full 3D image of FSs for this cube in the Fig.~\ref{fig:appendix:fs_threshold}. Cyan and red arrows specify DMFS and GFS filaments which do not show counterparts in another skeleton. The black dots show the distribution of dark matter particles, while yellow dots mark the position of galaxies with $M_{\ast} > 10^{9} M_{\odot}.$}
            \label{fig:bias_example}
        \end{figure}

        One important result of our analysis is that regardless of the  galaxy stellar mass threshold adopted, about 41\%~(Sec.~\ref{sec:different_tracers_mass}) of the dark matter filaments do not contain enough galaxies to be detected when using galaxies as tracers.  An example of this case is shown in  panel A of Fig.~\ref{fig:bias_example}. One possible explanation  is that there was not enough gas in these filaments to form enough galaxies of a given mass to allow the detection of a galaxy filament, given our adopted parameters in DisPerSE (Sec.~\ref{sec:disperse_settings}). 
         {Another possible explanation can be related to the filaments extraction. For instance,}
        lowering the adopted thresholds  helps to detect some of these `missing' filaments (panels B and C of Fig.~\ref{fig:bias_example}), as well as to detect additional `faint' filaments. Nevertheless,  even adopting a low-threshold~(panel C), some dark filaments  do not have counterparts in GFS  (and vice versa). We see this as a naturale reflection of the existence of the galaxy bias. We refer to Appendix~\ref{app:diff_tresh} for a more in depth analysis of the influence of adopted thresholds on the skeleton determination in 3D.  {Thus,   Fig.~\ref{fig:bias_example} shows that `missisng` filaments  are a consequence of  both  physical causes, and "operational" problems related to the definition and identification of filaments. The contribution of each cause cannot be reduced by fine tuning  the other. }

        We note that the FS also depend, to some extent, on the properties of the galaxy sample used as tracers (e.g. depending on the galaxy stellar mass limit adopted). This is expected, given that galaxy bias depends on galaxy properties~\citep[e.g.][]{Li+2006}. Our conclusions are in agreement with the work by \cite{Laigle+2018}, who  compare the COSMOS2015~\citep{Laigle+2016} galaxy sample (photometric and spectroscopic data) and the hydrodynamic simulations HORIZON-AGN to investigate the difference between skeletons obtained by  spectroscopic data, photometric data and corresponding dark matter data  from HORIZON-AGN. 

    \subsection{Virgo- and Coma-like clusters}
        Our selection of volumes centred on Virgo- and Coma-like haloes is motivated by the fact that there are good observational datasets available for these clusters in our local Universe. While we defer to a future work a more careful comparison between available data and predictions from the theoretical model that we have used in our study, we have carried out some basic comparisons with published results. 
        
        The size of our cubes (70 Mpc/h $\gg R_{vir} \sim 2$ Mpc/h) allows us to explore not only the region around these clusters, but the entire filamentary structure that is expected to flow into these clusters. As discussed in Sec.~\ref{sec:coincidence}, we do not find a better coincidence between DMFS and GFS in cubes containing Virgo- or Coma-like haloes. In general, this is due to the fact that the volume considered is 30-35 times larger than the size of the haloes. However, on smaller scales (for example, as inside cubes with a side of 14 Mpc/h as in the Fig.~\ref{fig:fss_examples}), the coincidence of the DMFS and GFS~($M_{\ast} > 10^{9} M_{\odot}$) near Virgo- and Coma-like improves to $73\pm10\%$ and $78\pm7\%$.  Accounting for low mass galaxies~($M_{\ast} < 10^{9} M_{\odot}$) does not significantly affect the results. Thus, the recovery of the cosmic web from galaxies around massive halos is better than for regions of more average density in the Universe.
        \par
        With our adopted parameters and in the typical regions we have considered, we find on average 3.3 (from 1 to 7) dark matter filaments for Coma-like haloes  and 2.8 for Virgo-like haloes (between 1 and 4).
        When considering the corresponding FS traced by galaxies with $M_{\ast} > 10^{9} M_{\odot}$, we find  2.6 filaments for Coma-like haloes (between is 1 and 4) and 1.9 filaments for Virgo-like haloes (between 1 and 3). Our results are consistent with an analysis based on SDSS7 data in a region around the Coma cluster of size similar to that of the volumes we have considered: \cite{Malavasi+2019} estimate a  median connectivity for Coma cluster of about 2.5. Approximately 1 dark matter filament going into clusters like Virgo and Coma cannot be traced by the distribution of galaxies more massive than  $10^9 M_{\odot}$. 
        \par
        The results we obtain for Virgo- and Coma-like haloes depend significantly on the criteria adopted to detect filaments: we have chosen a large scale and a high level of persistence, which leaves us with only the `strongest' (highest contrast) filaments. This should be taken into account when comparing with other studies. For example,~\cite{Castignani+2022_2, Castignani+2022_1} found 13 filaments around the Virgo cluster. By increasing lowering the threshold we adopt for filaments identification, as for example in panel C~(Fig.~\ref{fig:bias_example}), a larger number of small filaments can be found around Virgo-like clusters, bringing our results closer to those by~\cite{Castignani+2022_2}. 

  To summarize, the work presented in this paper shows that while the filament extraction strongly depends on the tracers used, and on the adoted parameters, the `strongest' filaments can be robustly identified.

\section{Data availability}
The data underlying this article will be shared on reasonable request to the corresponding author.
\section{Acknowledgment}
We thank the referee for the comments, that helped to improve the presentation of the results. 
We are indebted to Volker Springel for making the snapshots of the Millennium Simulation available to us.
DZ and BV acknowledge support from the INAF Mini Grant 2022 “Tracing filaments through cosmic time”  (PI Vulcani). The authors thank the hospitality of the International Space Science Institute (ISSI) in Bern (Switzerland). Regular group meetings in these institutes allowed the authors to make substantial progress on the project and finalize the present work. MH acknowledges funding from the Swiss National Science Foundation (SNF) via a PRIMA Grant PR00P2 193577 “From cosmic dawn to high noon: the role of black holes for young galaxies”.



\bibliographystyle{mnras}
\bibliography{main.bib} 

\begin{thebibliography}{}
\makeatletter
\relax
\def\mn@urlcharsother{\let\do\@makeother \do\$\do\&\do\#\do\^\do\_\do\%\do\~}
\def\mn@doi{\begingroup\mn@urlcharsother \@ifnextchar [ {\mn@doi@}
  {\mn@doi@[]}}
\def\mn@doi@[#1]#2{\def\@tempa{#1}\ifx\@tempa\@empty \href
  {http://dx.doi.org/#2} {doi:#2}\else \href {http://dx.doi.org/#2} {#1}\fi
  \endgroup}
\def\mn@eprint#1#2{\mn@eprint@#1:#2::\@nil}
\def\mn@eprint@arXiv#1{\href {http://arxiv.org/abs/#1} {{\tt arXiv:#1}}}
\def\mn@eprint@dblp#1{\href {http://dblp.uni-trier.de/rec/bibtex/#1.xml}
  {dblp:#1}}
\def\mn@eprint@#1:#2:#3:#4\@nil{\def\@tempa {#1}\def\@tempb {#2}\def\@tempc
  {#3}\ifx \@tempc \@empty \let \@tempc \@tempb \let \@tempb \@tempa \fi \ifx
  \@tempb \@empty \def\@tempb {arXiv}\fi \@ifundefined
  {mn@eprint@\@tempb}{\@tempb:\@tempc}{\expandafter \expandafter \csname
  mn@eprint@\@tempb\endcsname \expandafter{\@tempc}}}

\bibitem[\protect\citeauthoryear{{Alpaslan} et~al.,}{{Alpaslan}
  et~al.}{2014}]{Alpaslan+2014}
{Alpaslan} M.,  et~al., 2014, \mn@doi [\mnras] {10.1093/mnras/stt2136}, \href
  {https://ui.adsabs.harvard.edu/abs/2014MNRAS.438..177A} {438, 177}

\bibitem[\protect\citeauthoryear{{Arag{\'o}n-Calvo}, {van de Weygaert}, {Jones}
   \& {van der Hulst}}{{Arag{\'o}n-Calvo} et~al.}{2007}]{Aragon-Calvo+2007}
{Arag{\'o}n-Calvo} M.~A.,  {van de Weygaert} R.,  {Jones} B. J.~T.,   {van der
  Hulst} J.~M.,  2007, \mn@doi [\apjl] {10.1086/511633}, \href
  {https://ui.adsabs.harvard.edu/abs/2007ApJ...655L...5A} {655, L5}

\bibitem[\protect\citeauthoryear{{Arag{\'o}n-Calvo}, {van de Weygaert}  \&
  {Jones}}{{Arag{\'o}n-Calvo} et~al.}{2010}]{Aragon-Calvo+2010}
{Arag{\'o}n-Calvo} M.~A.,  {van de Weygaert} R.,   {Jones} B. J.~T.,  2010,
  \mn@doi [\mnras] {10.1111/j.1365-2966.2010.17263.x}, \href
  {https://ui.adsabs.harvard.edu/abs/2010MNRAS.408.2163A} {408, 2163}

\bibitem[\protect\citeauthoryear{{Barsanti} et~al.,}{{Barsanti}
  et~al.}{2022}]{Barsanti+2022}
{Barsanti} S.,  et~al., 2022, arXiv e-prints, \href
  {https://ui.adsabs.harvard.edu/abs/2022arXiv220810767B} {p. arXiv:2208.10767}

\bibitem[\protect\citeauthoryear{{Bermejo}, {Wilding}, {van de Weygaert},
  {Jones}, {Vegter}  \& {Efstathiou}}{{Bermejo} et~al.}{2022}]{Bermejo+2022}
{Bermejo} R.,  {Wilding} G.,  {van de Weygaert} R.,  {Jones} B. J.~T.,
  {Vegter} G.,   {Efstathiou} K.,  2022, arXiv e-prints, \href
  {https://ui.adsabs.harvard.edu/abs/2022arXiv220614655B} {p. arXiv:2206.14655}

\bibitem[\protect\citeauthoryear{{Beygu}, {Peletier}, {van der Hulst},
  {Jarrett}, {Kreckel}, {van de Weygaert}, {van Gorkom}  \&
  {Aragon-Calvo}}{{Beygu} et~al.}{2017}]{Beygu+2017}
{Beygu} B.,  {Peletier} R.~F.,  {van der Hulst} J.~M.,  {Jarrett} T.~H.,
  {Kreckel} K.,  {van de Weygaert} R.,  {van Gorkom} J.~H.,   {Aragon-Calvo}
  M.~A.,  2017, \mn@doi [\mnras] {10.1093/mnras/stw2362}, \href
  {https://ui.adsabs.harvard.edu/abs/2017MNRAS.464..666B} {464, 666}

\bibitem[\protect\citeauthoryear{{Blue Bird} et~al.,}{{Blue Bird}
  et~al.}{2020}]{Blue_Bird+2020}
{Blue Bird} J.,  et~al., 2020, \mn@doi [\mnras] {10.1093/mnras/stz3357}, \href
  {https://ui.adsabs.harvard.edu/abs/2020MNRAS.492..153B} {492, 153}

\bibitem[\protect\citeauthoryear{{Bond}, {Kofman}  \& {Pogosyan}}{{Bond}
  et~al.}{1996}]{Bond+1996}
{Bond} J.~R.,  {Kofman} L.,   {Pogosyan} D.,  1996, \mn@doi [\nat]
  {10.1038/380603a0}, \href
  {https://ui.adsabs.harvard.edu/abs/1996Natur.380..603B} {380, 603}

\bibitem[\protect\citeauthoryear{{Bonjean}, {Aghanim}, {Douspis}, {Malavasi}
  \& {Tanimura}}{{Bonjean} et~al.}{2020}]{Bonjean+2020}
{Bonjean} V.,  {Aghanim} N.,  {Douspis} M.,  {Malavasi} N.,   {Tanimura} H.,
  2020, \mn@doi [\aap] {10.1051/0004-6361/201937313}, \href
  {https://ui.adsabs.harvard.edu/abs/2020A&A...638A..75B} {638, A75}

\bibitem[\protect\citeauthoryear{{Boselli} \& {Gavazzi}}{{Boselli} \&
  {Gavazzi}}{2006}]{Boselli+2006}
{Boselli} A.,  {Gavazzi} G.,  2006, \mn@doi [\pasp] {10.1086/500691}, \href
  {https://ui.adsabs.harvard.edu/abs/2006PASP..118..517B} {118, 517}

\bibitem[\protect\citeauthoryear{{Bryant} et~al.,}{{Bryant}
  et~al.}{2015}]{Bryant+2015}
{Bryant} J.~J.,  et~al., 2015, \mn@doi [\mnras] {10.1093/mnras/stu2635}, \href
  {https://ui.adsabs.harvard.edu/abs/2015MNRAS.447.2857B} {447, 2857}

\bibitem[\protect\citeauthoryear{{Carr{\'o}n Duque}, {Migliaccio}, {Marinucci}
  \& {Vittorio}}{{Carr{\'o}n Duque} et~al.}{2022}]{Duque+2022}
{Carr{\'o}n Duque} J.,  {Migliaccio} M.,  {Marinucci} D.,   {Vittorio} N.,
  2022, \mn@doi [\aap] {10.1051/0004-6361/202141538}, \href
  {https://ui.adsabs.harvard.edu/abs/2022A&A...659A.166C} {659, A166}

\bibitem[\protect\citeauthoryear{{Castignani} et~al.,}{{Castignani}
  et~al.}{2022a}]{Castignani+2022_2}
{Castignani} G.,  et~al., 2022a, \mn@doi [\apjs] {10.3847/1538-4365/ac45f7},
  \href {https://ui.adsabs.harvard.edu/abs/2022ApJS..259...43C} {259, 43}

\bibitem[\protect\citeauthoryear{{Castignani} et~al.,}{{Castignani}
  et~al.}{2022b}]{Castignani+2022_1}
{Castignani} G.,  et~al., 2022b, \mn@doi [\aap] {10.1051/0004-6361/202040141},
  \href {https://ui.adsabs.harvard.edu/abs/2022A&A...657A...9C} {657, A9}

\bibitem[\protect\citeauthoryear{{Cautun}, {van de Weygaert}, {Jones}  \&
  {Frenk}}{{Cautun} et~al.}{2014}]{Cautun+2014}
{Cautun} M.,  {van de Weygaert} R.,  {Jones} B. J.~T.,   {Frenk} C.~S.,  2014,
  \mn@doi [\mnras] {10.1093/mnras/stu768}, \href
  {https://ui.adsabs.harvard.edu/abs/2014MNRAS.441.2923C} {441, 2923}

\bibitem[\protect\citeauthoryear{{Cautun}, {van de Weygaert}, {Jones}, {Frenk}
  \& {Hellwing}}{{Cautun} et~al.}{2015}]{Cautun+2015}
{Cautun} M.,  {van de Weygaert} R.,  {Jones} B. J.~T.,  {Frenk} C.~S.,
  {Hellwing} W.~A.,  2015, in Thirteenth Marcel Grossmann Meeting: On Recent
  Developments in Theoretical and Experimental General Relativity, Astrophysics
  and Relativistic Field Theories. pp 2115--2117 (\mn@eprint {arXiv}
  {1211.3574}), \mn@doi{10.1142/9789814623995_0371}

\bibitem[\protect\citeauthoryear{{Chang}, {Fang}, {Gu}, {Lin}, {Lu}  \&
  {Kong}}{{Chang} et~al.}{2022}]{Chang+2022}
{Chang} W.,  {Fang} G.,  {Gu} Y.,  {Lin} Z.,  {Lu} S.,   {Kong} X.,  2022,
  \mn@doi [\apj] {10.3847/1538-4357/ac8748}, \href
  {https://ui.adsabs.harvard.edu/abs/2022ApJ...936...47C} {936, 47}

\bibitem[\protect\citeauthoryear{{Chen}, {Ho}, {Freeman}, {Genovese}  \&
  {Wasserman}}{{Chen} et~al.}{2015}]{Chen+2015}
{Chen} Y.-C.,  {Ho} S.,  {Freeman} P.~E.,  {Genovese} C.~R.,   {Wasserman} L.,
  2015, \mn@doi [\mnras] {10.1093/mnras/stv1996}, \href
  {https://ui.adsabs.harvard.edu/abs/2015MNRAS.454.1140C} {454, 1140}

\bibitem[\protect\citeauthoryear{{Codis}, {Pogosyan}  \& {Pichon}}{{Codis}
  et~al.}{2018}]{Codis+2018}
{Codis} S.,  {Pogosyan} D.,   {Pichon} C.,  2018, \mn@doi [\mnras]
  {10.1093/mnras/sty1643}, \href
  {https://ui.adsabs.harvard.edu/abs/2018MNRAS.479..973C} {479, 973}

\bibitem[\protect\citeauthoryear{{Cornwell} et~al.,}{{Cornwell}
  et~al.}{2022}]{Cornwell+2022}
{Cornwell} D.~J.,  et~al., 2022, \mn@doi [\mnras] {10.1093/mnras/stac2777},
  \href {https://ui.adsabs.harvard.edu/abs/2022MNRAS.517.1678C} {517, 1678}

\bibitem[\protect\citeauthoryear{{Crone Odekon}, {Hallenbeck}, {Haynes},
  {Koopmann}, {Phi}  \& {Wolfe}}{{Crone Odekon} et~al.}{2018}]{Odekon+2018}
{Crone Odekon} M.,  {Hallenbeck} G.,  {Haynes} M.~P.,  {Koopmann} R.~A.,  {Phi}
  A.,   {Wolfe} P.-F.,  2018, \mn@doi [\apj] {10.3847/1538-4357/aaa1e8}, \href
  {https://ui.adsabs.harvard.edu/abs/2018ApJ...852..142C} {852, 142}

\bibitem[\protect\citeauthoryear{{DESI Collaboration} et~al.,}{{DESI
  Collaboration} et~al.}{2016}]{DESI+2016}
{DESI Collaboration} et~al., 2016, \mn@doi [arXiv e-prints]
  {10.48550/arXiv.1611.00036}, \href
  {https://ui.adsabs.harvard.edu/abs/2016arXiv161100036D} {p. arXiv:1611.00036}

\bibitem[\protect\citeauthoryear{{Darragh Ford} et~al.,}{{Darragh Ford}
  et~al.}{2019}]{Darragh_Ford+2019}
{Darragh Ford} E.,  et~al., 2019, \mn@doi [\mnras] {10.1093/mnras/stz2490},
  \href {https://ui.adsabs.harvard.edu/abs/2019MNRAS.489.5695D} {489, 5695}

\bibitem[\protect\citeauthoryear{{De Lucia}, {Boylan-Kolchin}, {Benson},
  {Fontanot}  \& {Monaco}}{{De Lucia} et~al.}{2010}]{DeLucia+2010}
{De Lucia} G.,  {Boylan-Kolchin} M.,  {Benson} A.~J.,  {Fontanot} F.,
  {Monaco} P.,  2010, \mn@doi [\mnras] {10.1111/j.1365-2966.2010.16806.x},
  \href {https://ui.adsabs.harvard.edu/abs/2010MNRAS.406.1533D} {406, 1533}

\bibitem[\protect\citeauthoryear{{De Lucia}, {Weinmann}, {Poggianti},
  {Arag{\'o}n-Salamanca}  \& {Zaritsky}}{{De Lucia}
  et~al.}{2012}]{De_Lucia+2012}
{De Lucia} G.,  {Weinmann} S.,  {Poggianti} B.~M.,  {Arag{\'o}n-Salamanca} A.,
   {Zaritsky} D.,  2012, \mn@doi [\mnras] {10.1111/j.1365-2966.2012.20983.x},
  \href {https://ui.adsabs.harvard.edu/abs/2012MNRAS.423.1277D} {423, 1277}

\bibitem[\protect\citeauthoryear{{De Lucia}, {Tornatore}, {Frenk}, {Helmi},
  {Navarro}  \& {White}}{{De Lucia} et~al.}{2014}]{DeLucia+2014}
{De Lucia} G.,  {Tornatore} L.,  {Frenk} C.~S.,  {Helmi} A.,  {Navarro} J.~F.,
   {White} S. D.~M.,  2014, \mn@doi [\mnras] {10.1093/mnras/stu1752}, \href
  {https://ui.adsabs.harvard.edu/abs/2014MNRAS.445..970D} {445, 970}

\bibitem[\protect\citeauthoryear{{Donnari}, {Pillepich}, {Nelson}, {Marinacci},
  {Vogelsberger}  \& {Hernquist}}{{Donnari} et~al.}{2021}]{Donnari+2021}
{Donnari} M.,  {Pillepich} A.,  {Nelson} D.,  {Marinacci} F.,  {Vogelsberger}
  M.,   {Hernquist} L.,  2021, \mn@doi [\mnras] {10.1093/mnras/stab1950}, \href
  {https://ui.adsabs.harvard.edu/abs/2021MNRAS.506.4760D} {506, 4760}

\bibitem[\protect\citeauthoryear{{Dressler}}{{Dressler}}{1980}]{Dressler+1980_red}
{Dressler} A.,  1980, \mn@doi [\apjs] {10.1086/190663}, \href
  {https://ui.adsabs.harvard.edu/abs/1980ApJS...42..565D} {42, 565}

\bibitem[\protect\citeauthoryear{{Dubois} et~al.,}{{Dubois}
  et~al.}{2014}]{Dubois+2014}
{Dubois} Y.,  et~al., 2014, \mn@doi [\mnras] {10.1093/mnras/stu1227}, \href
  {https://ui.adsabs.harvard.edu/abs/2014MNRAS.444.1453D} {444, 1453}

\bibitem[\protect\citeauthoryear{{Einasto}, {Kipper}, {Tenjes}, {Einasto},
  {Tempel}  \& {Liivam{\"a}gi}}{{Einasto} et~al.}{2022}]{Einasto+2022}
{Einasto} M.,  {Kipper} R.,  {Tenjes} P.,  {Einasto} J.,  {Tempel} E.,
  {Liivam{\"a}gi} L.~J.,  2022, arXiv e-prints, \href
  {https://ui.adsabs.harvard.edu/abs/2022arXiv221010761E} {p. arXiv:2210.10761}

\bibitem[\protect\citeauthoryear{{Gal{\'a}rraga-Espinosa}, {Aghanim}, {Langer},
  {Gouin}  \& {Malavasi}}{{Gal{\'a}rraga-Espinosa}
  et~al.}{2020}]{Galarraga-Espinosa+2020}
{Gal{\'a}rraga-Espinosa} D.,  {Aghanim} N.,  {Langer} M.,  {Gouin} C.,
  {Malavasi} N.,  2020, \mn@doi [\aap] {10.1051/0004-6361/202037986}, \href
  {https://ui.adsabs.harvard.edu/abs/2020A&A...641A.173G} {641, A173}

\bibitem[\protect\citeauthoryear{{Gouin}, {Bonnaire}  \& {Aghanim}}{{Gouin}
  et~al.}{2021}]{Gouin+2021}
{Gouin} C.,  {Bonnaire} T.,   {Aghanim} N.,  2021, \mn@doi [\aap]
  {10.1051/0004-6361/202140327}, \href
  {https://ui.adsabs.harvard.edu/abs/2021A&A...651A..56G} {651, A56}

\bibitem[\protect\citeauthoryear{{Groener}, {Goldberg}  \& {Sereno}}{{Groener}
  et~al.}{2016}]{Groener+2016}
{Groener} A.~M.,  {Goldberg} D.~M.,   {Sereno} M.,  2016, \mn@doi [\mnras]
  {10.1093/mnras/stv2341}, \href
  {https://ui.adsabs.harvard.edu/abs/2016MNRAS.455..892G} {455, 892}

\bibitem[\protect\citeauthoryear{{Guo}, {Tempel}  \& {Libeskind}}{{Guo}
  et~al.}{2015}]{Guo+2015}
{Guo} Q.,  {Tempel} E.,   {Libeskind} N.~I.,  2015, \mn@doi [\apj]
  {10.1088/0004-637X/800/2/112}, \href
  {https://ui.adsabs.harvard.edu/abs/2015ApJ...800..112G} {800, 112}

\bibitem[\protect\citeauthoryear{{Haynes}}{{Haynes}}{1985}]{Haynes+1985}
{Haynes} M.~P.,  1985, in European Southern Observatory Conference and Workshop
  Proceedings. pp 45--50

\bibitem[\protect\citeauthoryear{{Hirschmann}, {De Lucia}  \&
  {Fontanot}}{{Hirschmann} et~al.}{2016}]{Hirschman+2016}
{Hirschmann} M.,  {De Lucia} G.,   {Fontanot} F.,  2016, \mn@doi [\mnras]
  {10.1093/mnras/stw1318}, \href
  {https://ui.adsabs.harvard.edu/abs/2016MNRAS.461.1760H} {461, 1760}

\bibitem[\protect\citeauthoryear{{Hoyle}, {Vogeley}  \& {Rojas}}{{Hoyle}
  et~al.}{2005}]{Hoyle+2005}
{Hoyle} F.,  {Vogeley} M.~S.,   {Rojas} R.~R.,  2005, in American Astronomical
  Society Meeting Abstracts \#206. p. 10.02

\bibitem[\protect\citeauthoryear{{Huchra} et~al.,}{{Huchra}
  et~al.}{2012}]{Huchra+2012}
{Huchra} J.~P.,  et~al., 2012, \mn@doi [\apjs] {10.1088/0067-0049/199/2/26},
  \href {https://ui.adsabs.harvard.edu/abs/2012ApJS..199...26H} {199, 26}

\bibitem[\protect\citeauthoryear{{Inoue}, {Si}, {Okamoto}  \&
  {Nishigaki}}{{Inoue} et~al.}{2022}]{Inoue+2022}
{Inoue} S.,  {Si} X.,  {Okamoto} T.,   {Nishigaki} M.,  2022, \mn@doi [\mnras]
  {10.1093/mnras/stac2055}, \href
  {https://ui.adsabs.harvard.edu/abs/2022MNRAS.515.4065I} {515, 4065}

\bibitem[\protect\citeauthoryear{{Jeong}, {Desjacques}  \& {Schmidt}}{{Jeong}
  et~al.}{2018}]{Desjacques+2018}
{Jeong} D.,  {Desjacques} V.,   {Schmidt} F.,  2018, in American Astronomical
  Society Meeting Abstracts \#231. p. 420.03

\bibitem[\protect\citeauthoryear{{Kaiser}}{{Kaiser}}{1984}]{Kaiser+1984}
{Kaiser} N.,  1984, \mn@doi [\apjl] {10.1086/184341}, \href
  {https://ui.adsabs.harvard.edu/abs/1984ApJ...284L...9K} {284, L9}

\bibitem[\protect\citeauthoryear{{Kim} et~al.,}{{Kim} et~al.}{2016}]{Kim+2016}
{Kim} S.,  et~al., 2016, \mn@doi [\apj] {10.3847/1538-4357/833/2/207}, \href
  {https://ui.adsabs.harvard.edu/abs/2016ApJ...833..207K} {833, 207}

\bibitem[\protect\citeauthoryear{{Kleiner}, {Pimbblet}, {Jones}, {Koribalski}
  \& {Serra}}{{Kleiner} et~al.}{2017}]{Kleiner+2017}
{Kleiner} D.,  {Pimbblet} K.~A.,  {Jones} D.~H.,  {Koribalski} B.~S.,   {Serra}
  P.,  2017, \mn@doi [\mnras] {10.1093/mnras/stw3328}, \href
  {https://ui.adsabs.harvard.edu/abs/2017MNRAS.466.4692K} {466, 4692}

\bibitem[\protect\citeauthoryear{{Knebe}, {Gill}, {Gibson}, {Lewis}, {Ibata}
  \& {Dopita}}{{Knebe} et~al.}{2004}]{Knebe+2004}
{Knebe} A.,  {Gill} S. P.~D.,  {Gibson} B.~K.,  {Lewis} G.~F.,  {Ibata} R.~A.,
   {Dopita} M.~A.,  2004, \mn@doi [\apj] {10.1086/381306}, \href
  {https://ui.adsabs.harvard.edu/abs/2004ApJ...603....7K} {603, 7}

\bibitem[\protect\citeauthoryear{Kraljic et~al.,}{Kraljic
  et~al.}{2017}]{Kraljic+2018}
Kraljic K.,  et~al., 2017, \mn@doi [Monthly Notices of the Royal Astronomical
  Society] {10.1093/mnras/stx2638}, 474, 547

\bibitem[\protect\citeauthoryear{{Kraljic} et~al.,}{{Kraljic}
  et~al.}{2019}]{Kraljic+2019}
{Kraljic} K.,  et~al., 2019, \mn@doi [\mnras] {10.1093/mnras/sty3216}, \href
  {https://ui.adsabs.harvard.edu/abs/2019MNRAS.483.3227K} {483, 3227}

\bibitem[\protect\citeauthoryear{{Kreckel}, {Peebles}, {van Gorkom}, {van de
  Weygaert}  \& {van der Hulst}}{{Kreckel} et~al.}{2011}]{Kreckel+2011}
{Kreckel} K.,  {Peebles} P.~J.~E.,  {van Gorkom} J.~H.,  {van de Weygaert} R.,
   {van der Hulst} J.~M.,  2011, \mn@doi [\aj] {10.1088/0004-6256/141/6/204},
  \href {https://ui.adsabs.harvard.edu/abs/2011AJ....141..204K} {141, 204}

\bibitem[\protect\citeauthoryear{{Kuchner} et~al.,}{{Kuchner}
  et~al.}{2020}]{Kuchner+2020}
{Kuchner} U.,  et~al., 2020, \mn@doi [\mnras] {10.1093/mnras/staa1083}, \href
  {https://ui.adsabs.harvard.edu/abs/2020MNRAS.494.5473K} {494, 5473}

\bibitem[\protect\citeauthoryear{{Kuchner} et~al.,}{{Kuchner}
  et~al.}{2021}]{Kuchner+2021}
{Kuchner} U.,  et~al., 2021, \mn@doi [\mnras] {10.1093/mnras/stab567}, \href
  {https://ui.adsabs.harvard.edu/abs/2021MNRAS.503.2065K} {503, 2065}

\bibitem[\protect\citeauthoryear{{Kuutma}, {Tamm}  \& {Tempel}}{{Kuutma}
  et~al.}{2017}]{Kuutma+2017}
{Kuutma} T.,  {Tamm} A.,   {Tempel} E.,  2017, \mn@doi [\aap]
  {10.1051/0004-6361/201730526}, \href
  {https://ui.adsabs.harvard.edu/abs/2017A&A...600L...6K} {600, L6}

\bibitem[\protect\citeauthoryear{{Laigle} et~al.,}{{Laigle}
  et~al.}{2016}]{Laigle+2016}
{Laigle} C.,  et~al., 2016, \mn@doi [\apjs] {10.3847/0067-0049/224/2/24}, \href
  {https://ui.adsabs.harvard.edu/abs/2016ApJS..224...24L} {224, 24}

\bibitem[\protect\citeauthoryear{{Laigle} et~al.,}{{Laigle}
  et~al.}{2017}]{Laigle+2017}
{Laigle} C.,  et~al., 2017, VizieR Online Data Catalog, \href
  {https://ui.adsabs.harvard.edu/abs/2017yCat..22240024L} {p. J/ApJS/224/24}

\bibitem[\protect\citeauthoryear{{Laigle} et~al.,}{{Laigle}
  et~al.}{2018}]{Laigle+2018}
{Laigle} C.,  et~al., 2018, \mn@doi [\mnras] {10.1093/mnras/stx3055}, \href
  {https://ui.adsabs.harvard.edu/abs/2018MNRAS.474.5437L} {474, 5437}

\bibitem[\protect\citeauthoryear{{Lee}, {Kim}, {Rey}  \& {Chung}}{{Lee}
  et~al.}{2021}]{Lee+2021}
{Lee} Y.,  {Kim} S.,  {Rey} S.-C.,   {Chung} J.,  2021, \mn@doi [\apj]
  {10.3847/1538-4357/abcaa0}, \href
  {https://ui.adsabs.harvard.edu/abs/2021ApJ...906...68L} {906, 68}

\bibitem[\protect\citeauthoryear{{Li}, {Kauffmann}, {Jing}, {White},
  {B{\"o}rner}  \& {Cheng}}{{Li} et~al.}{2006}]{Li+2006}
{Li} C.,  {Kauffmann} G.,  {Jing} Y.~P.,  {White} S. D.~M.,  {B{\"o}rner} G.,
  {Cheng} F.~Z.,  2006, \mn@doi [\mnras] {10.1111/j.1365-2966.2006.10066.x},
  \href {https://ui.adsabs.harvard.edu/abs/2006MNRAS.368...21L} {368, 21}

\bibitem[\protect\citeauthoryear{{Libeskind} et~al.,}{{Libeskind}
  et~al.}{2018}]{Libeskind+2018}
{Libeskind} N.~I.,  et~al., 2018, \mn@doi [\mnras] {10.1093/mnras/stx1976},
  \href {https://ui.adsabs.harvard.edu/abs/2018MNRAS.473.1195L} {473, 1195}

\bibitem[\protect\citeauthoryear{{Malavasi} et~al.,}{{Malavasi}
  et~al.}{2017}]{Malavasi+2017}
{Malavasi} N.,  et~al., 2017, \mn@doi [\mnras] {10.1093/mnras/stw2864}, \href
  {https://ui.adsabs.harvard.edu/abs/2017MNRAS.465.3817M} {465, 3817}

\bibitem[\protect\citeauthoryear{{Malavasi}, {Aghanim}, {Tanimura}, {Bonjean}
  \& {Douspis}}{{Malavasi} et~al.}{2020a}]{Malavasi+2019}
{Malavasi} N.,  {Aghanim} N.,  {Tanimura} H.,  {Bonjean} V.,   {Douspis} M.,
  2020a, \mn@doi [\aap] {10.1051/0004-6361/201936629}, \href
  {https://ui.adsabs.harvard.edu/abs/2020A&A...634A..30M} {634, A30}

\bibitem[\protect\citeauthoryear{{Malavasi}, {Aghanim}, {Douspis}, {Tanimura}
  \& {Bonjean}}{{Malavasi} et~al.}{2020b}]{Malavasi+2020}
{Malavasi} N.,  {Aghanim} N.,  {Douspis} M.,  {Tanimura} H.,   {Bonjean} V.,
  2020b, \mn@doi [\aap] {10.1051/0004-6361/202037647}, \href
  {https://ui.adsabs.harvard.edu/abs/2020A&A...642A..19M} {642, A19}

\bibitem[\protect\citeauthoryear{{Okabe}, {Futamase}, {Kajisawa}  \&
  {Kuroshima}}{{Okabe} et~al.}{2014}]{Okabe+2014}
{Okabe} N.,  {Futamase} T.,  {Kajisawa} M.,   {Kuroshima} R.,  2014, \mn@doi
  [\apj] {10.1088/0004-637X/784/2/90}, \href
  {https://ui.adsabs.harvard.edu/abs/2014ApJ...784...90O} {784, 90}

\bibitem[\protect\citeauthoryear{{Paccagnella} et~al.,}{{Paccagnella}
  et~al.}{2016}]{Paccagnella+2016}
{Paccagnella} A.,  et~al., 2016, \mn@doi [\apjl] {10.3847/2041-8205/816/2/L25},
  \href {https://ui.adsabs.harvard.edu/abs/2016ApJ...816L..25P} {816, L25}

\bibitem[\protect\citeauthoryear{{Park} et~al.,}{{Park}
  et~al.}{2022}]{Park+2022}
{Park} M.,  et~al., 2022, \mn@doi [\mnras] {10.1093/mnras/stac1773}, \href
  {https://ui.adsabs.harvard.edu/abs/2022MNRAS.515..213P} {515, 213}

\bibitem[\protect\citeauthoryear{{Peebles}}{{Peebles}}{1980}]{Peebles_1980_book}
{Peebles} P.~J.~E.,  1980, {The large-scale structure of the universe}

\bibitem[\protect\citeauthoryear{{Pfeifer}, {Libeskind}, {Hoffman}, {Hellwing},
  {Bilicki}  \& {Naidoo}}{{Pfeifer} et~al.}{2022}]{Pfeife+2022}
{Pfeifer} S.,  {Libeskind} N.~I.,  {Hoffman} Y.,  {Hellwing} W.~A.,  {Bilicki}
  M.,   {Naidoo} K.,  2022, \mn@doi [\mnras] {10.1093/mnras/stac1382}, \href
  {https://ui.adsabs.harvard.edu/abs/2022MNRAS.514..470P} {514, 470}

\bibitem[\protect\citeauthoryear{{Poggianti} et~al.,}{{Poggianti}
  et~al.}{2006}]{Poggianti+2006}
{Poggianti} B.~M.,  et~al., 2006, \mn@doi [\apj] {10.1086/500666}, \href
  {https://ui.adsabs.harvard.edu/abs/2006ApJ...642..188P} {642, 188}

\bibitem[\protect\citeauthoryear{{Rojas}, {Vogeley}, {Hoyle}  \&
  {Brinkmann}}{{Rojas} et~al.}{2004}]{Rojas+2004}
{Rojas} R.~R.,  {Vogeley} M.~S.,  {Hoyle} F.,   {Brinkmann} J.,  2004, \mn@doi
  [\apj] {10.1086/425225}, \href
  {https://ui.adsabs.harvard.edu/abs/2004ApJ...617...50R} {617, 50}

\bibitem[\protect\citeauthoryear{{Rost} et~al.,}{{Rost}
  et~al.}{2021}]{Rost+2021}
{Rost} A.,  et~al., 2021, \mn@doi [\mnras] {10.1093/mnras/staa3792}, \href
  {https://ui.adsabs.harvard.edu/abs/2021MNRAS.502..714R} {502, 714}

\bibitem[\protect\citeauthoryear{{Sarron}, {Adami}, {Durret}  \&
  {Laigle}}{{Sarron} et~al.}{2019}]{Sarron+2019}
{Sarron} F.,  {Adami} C.,  {Durret} F.,   {Laigle} C.,  2019, \mn@doi [\aap]
  {10.1051/0004-6361/201935394}, \href
  {https://ui.adsabs.harvard.edu/abs/2019A&A...632A..49S} {632, A49}

\bibitem[\protect\citeauthoryear{{Scoccimarro}}{{Scoccimarro}}{2000}]{Scoccimarro+2000}
{Scoccimarro} R.,  2000, \mn@doi [\apj] {10.1086/317248}, \href
  {https://ui.adsabs.harvard.edu/abs/2000ApJ...544..597S} {544, 597}

\bibitem[\protect\citeauthoryear{{Singh}, {Mahajan}  \& {Bagla}}{{Singh}
  et~al.}{2020}]{Singh+2020}
{Singh} A.,  {Mahajan} S.,   {Bagla} J.~S.,  2020, \mn@doi [\mnras]
  {10.1093/mnras/staa1913}, \href
  {https://ui.adsabs.harvard.edu/abs/2020MNRAS.497.2265S} {497, 2265}

\bibitem[\protect\citeauthoryear{{Skrutskie} et~al.,}{{Skrutskie}
  et~al.}{2006}]{Skrutskie+2006}
{Skrutskie} M.~F.,  et~al., 2006, \mn@doi [\aj] {10.1086/498708}, \href
  {https://ui.adsabs.harvard.edu/abs/2006AJ....131.1163S} {131, 1163}

\bibitem[\protect\citeauthoryear{{Song} et~al.,}{{Song}
  et~al.}{2021}]{Song+2021}
{Song} H.,  et~al., 2021, \mn@doi [\mnras] {10.1093/mnras/staa3981}, \href
  {https://ui.adsabs.harvard.edu/abs/2021MNRAS.501.4635S} {501, 4635}

\bibitem[\protect\citeauthoryear{{Sousbie}}{{Sousbie}}{2011}]{Sousbie+2011}
{Sousbie} T.,  2011, \mn@doi [\mnras] {10.1111/j.1365-2966.2011.18394.x}, \href
  {https://ui.adsabs.harvard.edu/abs/2011MNRAS.414..350S} {414, 350}

\bibitem[\protect\citeauthoryear{{Sousbie}, {Pichon}  \& {Kawahara}}{{Sousbie}
  et~al.}{2011}]{Sousbie_etal+2011}
{Sousbie} T.,  {Pichon} C.,   {Kawahara} H.,  2011, \mn@doi [\mnras]
  {10.1111/j.1365-2966.2011.18395.x}, \href
  {https://ui.adsabs.harvard.edu/abs/2011MNRAS.414..384S} {414, 384}

\bibitem[\protect\citeauthoryear{{Springel} et~al.,}{{Springel}
  et~al.}{2005}]{Springel+2005}
{Springel} V.,  et~al., 2005, \mn@doi [\nat] {10.1038/nature03597}, \href
  {https://ui.adsabs.harvard.edu/abs/2005Natur.435..629S} {435, 629}

\bibitem[\protect\citeauthoryear{{Taylor}}{{Taylor}}{2020}]{Taylor+2020}
{Taylor} E.,  2020, in The Build-Up of Galaxies through Multiple Tracers and
  Facilities. p.~75, \mn@doi{10.5281/zenodo.3756572}

\bibitem[\protect\citeauthoryear{{Tegmark} \& {Peebles}}{{Tegmark} \&
  {Peebles}}{1998}]{Tegmark+1998}
{Tegmark} M.,  {Peebles} P.~J.~E.,  1998, \mn@doi [\apjl] {10.1086/311426},
  \href {https://ui.adsabs.harvard.edu/abs/1998ApJ...500L..79T} {500, L79}

\bibitem[\protect\citeauthoryear{{Tempel}, {Stoica}, {Mart{\'\i}nez},
  {Liivam{\"a}gi}, {Castellan}  \& {Saar}}{{Tempel} et~al.}{2014}]{Tempel+2014}
{Tempel} E.,  {Stoica} R.~S.,  {Mart{\'\i}nez} V.~J.,  {Liivam{\"a}gi} L.~J.,
  {Castellan} G.,   {Saar} E.,  2014, \mn@doi [\mnras] {10.1093/mnras/stt2454},
  \href {https://ui.adsabs.harvard.edu/abs/2014MNRAS.438.3465T} {438, 3465}

\bibitem[\protect\citeauthoryear{{Tully}}{{Tully}}{1982}]{Tully+1982}
{Tully} R.~B.,  1982, \mn@doi [\apj] {10.1086/159999}, \href
  {https://ui.adsabs.harvard.edu/abs/1982ApJ...257..389T} {257, 389}

\bibitem[\protect\citeauthoryear{{Vulcani}, {Poggianti}, {Finn}, {Rudnick},
  {Desai}  \& {Bamford}}{{Vulcani} et~al.}{2010}]{Vulcani+2010}
{Vulcani} B.,  {Poggianti} B.~M.,  {Finn} R.~A.,  {Rudnick} G.,  {Desai} V.,
  {Bamford} S.,  2010, \mn@doi [\apjl] {10.1088/2041-8205/710/1/L1}, \href
  {https://ui.adsabs.harvard.edu/abs/2010ApJ...710L...1V} {710, L1}

\bibitem[\protect\citeauthoryear{{Vulcani} et~al.,}{{Vulcani}
  et~al.}{2019}]{Vulcani+2019}
{Vulcani} B.,  et~al., 2019, \mn@doi [\mnras] {10.1093/mnras/stz1399}, \href
  {https://ui.adsabs.harvard.edu/abs/2019MNRAS.487.2278V} {487, 2278}

\bibitem[\protect\citeauthoryear{{Xie}, {De Lucia}, {Hirschmann}, {Fontanot}
  \& {Zoldan}}{{Xie} et~al.}{2017}]{Xie+2017}
{Xie} L.,  {De Lucia} G.,  {Hirschmann} M.,  {Fontanot} F.,   {Zoldan} A.,
  2017, \mn@doi [\mnras] {10.1093/mnras/stx889}, \href
  {https://ui.adsabs.harvard.edu/abs/2017MNRAS.469..968X} {469, 968}

\bibitem[\protect\citeauthoryear{{Xie}, {De Lucia}, {Hirschmann}  \&
  {Fontanot}}{{Xie} et~al.}{2020}]{Xie+2020}
{Xie} L.,  {De Lucia} G.,  {Hirschmann} M.,   {Fontanot} F.,  2020, \mn@doi
  [\mnras] {10.1093/mnras/staa2370}, \href
  {https://ui.adsabs.harvard.edu/abs/2020MNRAS.498.4327X} {498, 4327}

\bibitem[\protect\citeauthoryear{{van de Weygaert} \& {Schaap}}{{van de
  Weygaert} \& {Schaap}}{2009}]{Weygaert+2009}
{van de Weygaert} R.,  {Schaap} W.,  2009, in {Mart{\'\i}nez} V.~J.,  {Saar}
  E.,  {Mart{\'\i}nez-Gonz{\'a}lez} E.,   {Pons-Border{\'\i}a} M.~J.,  eds, ,
  Vol.~665, Data Analysis in Cosmology.
pp 291--413, \mn@doi{10.1007/978-3-540-44767-2_11}

\makeatother
\end{thebibliography}



\appendix
\section{The effect of choosing different persistence thresholds in DisPerSE}
\label{app:diff_tresh}

    As discussed in the Introduction, the filament identification is strongly linked to the adopted method and assumptions. The level of details in the resulting structure and the smallest filaments that are resolved can vary significantly from work to work.
    In DisPerSE, which is the tool used in the work, it is possible to tune parameters like the persistence threshold: all structures below the adopted value  are not taken into account for calculating the FS. In this appendix we explore how the results presented in the paper  depend on the choice of the threshold and how the overlap between DMFS and GFS depends on it. 
    \par
    For simplicity, we here consider only one cube and run DisPerSE lowering the threshold of the persistence. In the first run, we adopt values that are approximately half the values adopted in our work, in the second run we half again such values. Hence, we adopt values of  {$5 \cdot 10^{8}$} and  {$3 \cdot 10^{8}$} for the dark matter,  {$10^{3}$} and  {10} for galaxies, in the first and second runs, respectively. This choice ensure us to always  obtain DMFS and GFS of approximately the same total length. 
    \par
\begin{figure}
        \centering
        \includegraphics[width=1\linewidth]{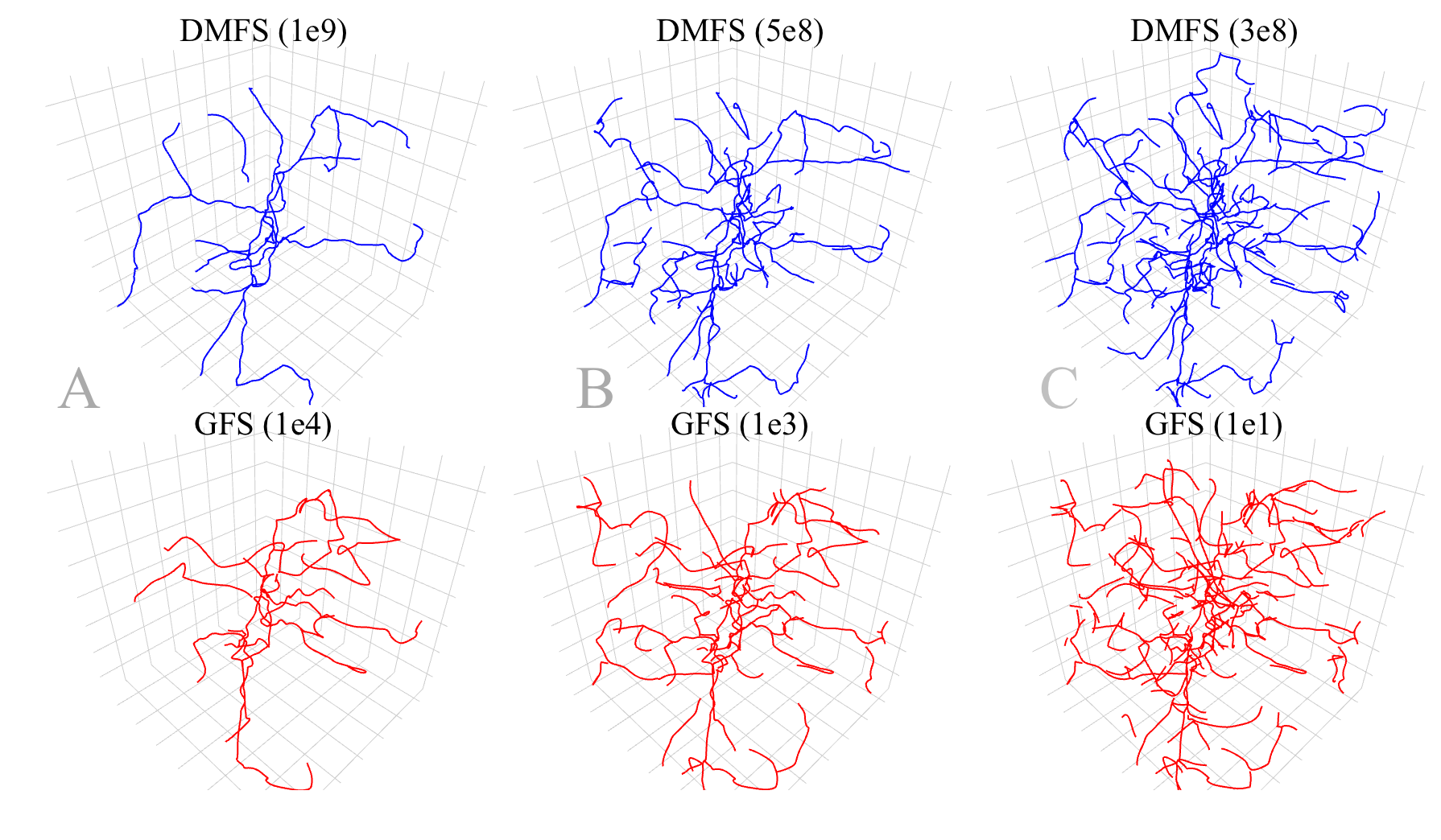}
       \caption{Variation of the threshold level for DMFS~(top) and GFS~(bottom) for one cube. The left column shows adopted threshold in main part of work. The exact values of the threshold are on the plot.  Skeletons are full version of slices In the Fig.~\ref{fig:bias_example}. }
       \label{fig:appendix:fs_threshold}
    \end{figure}
\begin{figure}
        \centering
        \includegraphics[width=1\linewidth]{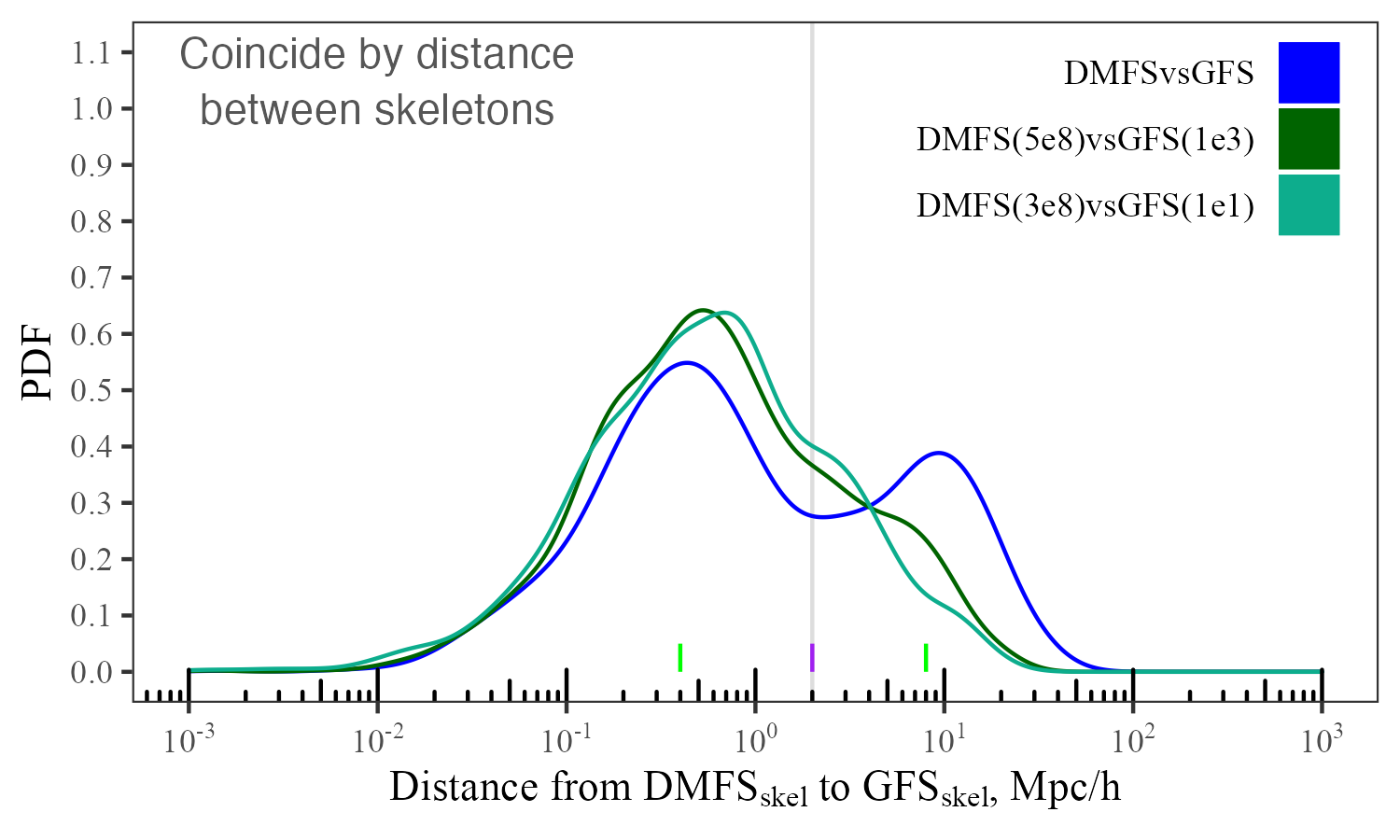}
       \caption{Coincidence by distance between DMFS and GFS for different thresholds. Thresholds corresponds to Fig.~\ref{fig:appendix_fs_threshold_coin} and Fig.~\ref{fig:bias_example}. Case of panel A is shown by blue color, case of panel B is shown by dark green color and case of panel C is shown by blue-green color.
       Notches were left from previous estimates on values 0.4, 2 and 8 Mpc/h. Separation by coincidence is also left at the value of 2 Mpc/h. Each line represents all cubes together.}
       \label{fig:appendix:fs_threshold_coin}
    \end{figure}

    Results are presented in Fig.~\ref{fig:appendix:fs_threshold}.   
    The DMFS and GFS used in our work are shown in the left column~(corresponding to panel A in  Fig.~\ref{fig:bias_example}). The central~(corresponding to panel B) and right~(panel C) columns show the consequences of lowering the thresholds of the persistence from  {$10^{9}$} to  {$5 \cdot 10^{8}$}( {$3 \cdot 10^{8}$}) for the DMFS and from  {$10^{4}$} to  {$10^{3}$}( {10}) for the GFS. 
    When decreasing the threshold level, new filaments appear in both the DMFS and GFS: as shown in Fig.~\ref{fig:bias_example}, some dark matter filaments acquire counterparts in GFS~(and vice versa). As a consequence, also  the coincidence by distance between skeletons improves~(Fig.~\ref{fig:appendix:fs_threshold_coin}) and goes from ~60\% (Sec.~\ref{sec:coincidence}) up to 76\% and 79\% respectively for panel B and panel C in Fig.~\ref{fig:appendix:fs_threshold_coin}. However, we note about 20 percent of dark matter filaments still do not have a GFS counterpart~(within a radius of 2 Mpc/h) in case of each selected thresholds. We associate the increase in  coincidence  not only with the actual increase in the matching of two skeletons, but also with the fact that two larger structures coincide more strongly than smaller structures by construction. To sum up, simply lowering the threshold does not lead to guaranteed coincidence of the skeletons, but rather adds a lot of new elements, some of which intersect.
   
    \par

    \section{Subsampling effects}
    \label{app:subsampling}

    The number of particles/galaxies in DM and galaxy cubes differs for each extracted cube by a factor $\sim$1000. In this Appendix, we  briefly discuss  the impact of having different sample sizes, by artificially reducing the number of DM particles in a cube. We have checked that the results presented below are valid when considering any of the 27 cubes we have at our disposal.
    \par
\begin{figure}
        \centering\includegraphics[width=1\linewidth]{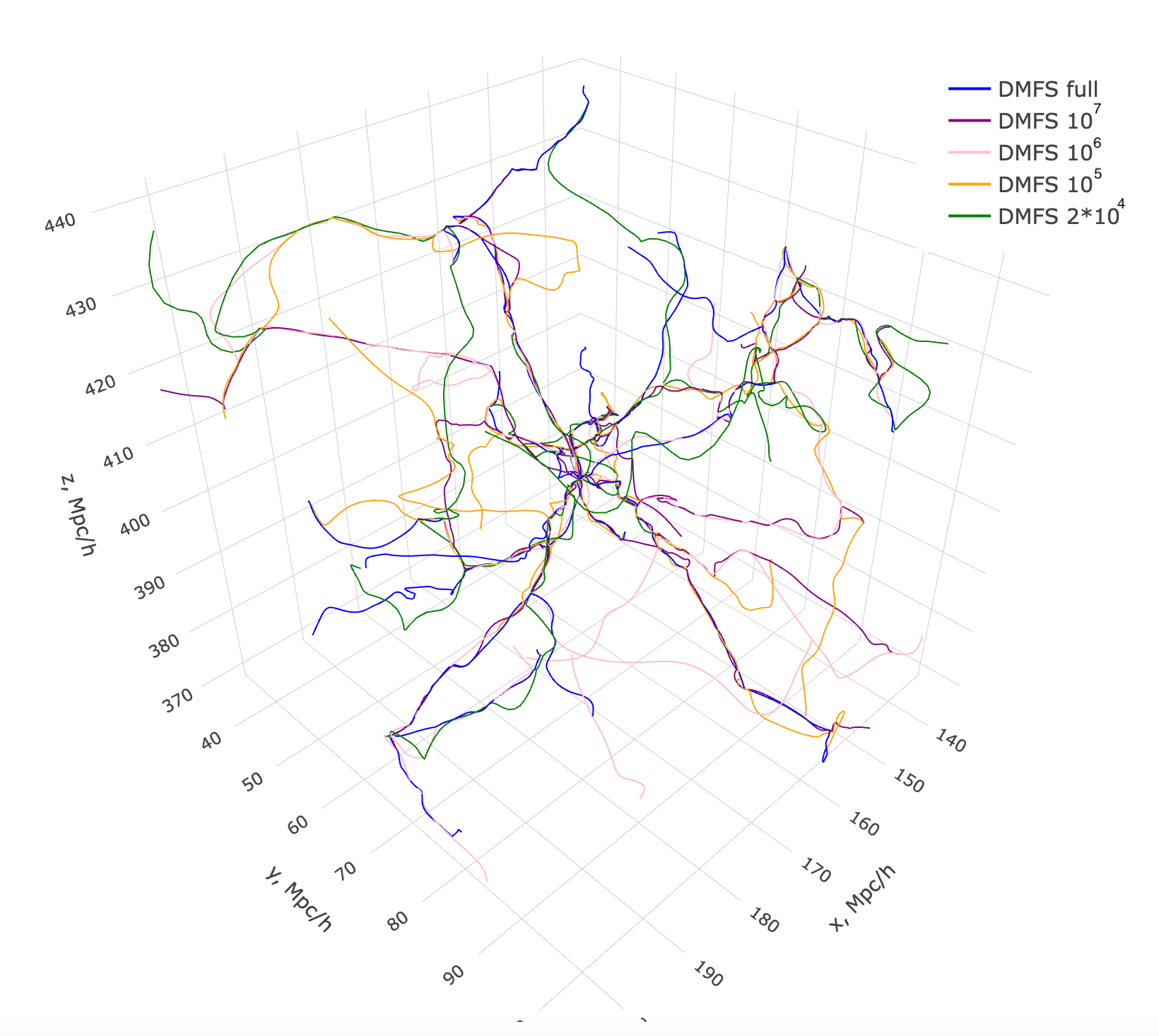}
        \caption{Dark matter filaments  extracted using the subsampled cubes, as explaiend in the text. The blue DMFS is extracted considering the  full DM cube ($\sim 3 \cdot 10^{7}$ particles) and  corresponds to the DMFS in the main bory of the paper, while the other lines show the DMFS extracted using the randomly subsampled cubes: purple:  $10^{7}$,  pink: $10^{6}$,  orange:$10^{5}$ and yellow: $2\cdot 10^{4}$ (corresponding to the typical number of galaxies over the same area). We omit the $5\cdot 10^{6}$ and $2\cdot 10^{4}$ DMFS clarity.}
      \label{fig:appendix:fs_subset_3d}
    \end{figure}

    We randomly extracted  from a cube containing ~$3 \cdot 10^{7}$ DM particles 6 different subcubes, each with a different number of particles:  $10^{7}$, $5\cdot 10^{6}$, $10^{6}$, $10^{5}$, $2\cdot 10^{4}$ and $10^{4}$. The two last cases mimic the typical number of galaxies with $M_{\ast} > 10^{9} M_{\odot}$
    over a similar region. 
    For each sub-sample, we extracted the DMFS, as explained in  Sec.~\ref{sec:data_and_methods}, fine tuning the parameters to obtain a similar filament length in all realisations. Fig.~\ref{fig:appendix:fs_subset_3d} shows 5 of 7 extracted FS\footnote{We excluded two FS~($5\cdot 10^{6}$ and $2\cdot 10^{4}$) for sake of clarity in the plot.}.

\begin{figure}
        \centering
        \includegraphics[width=1\linewidth]{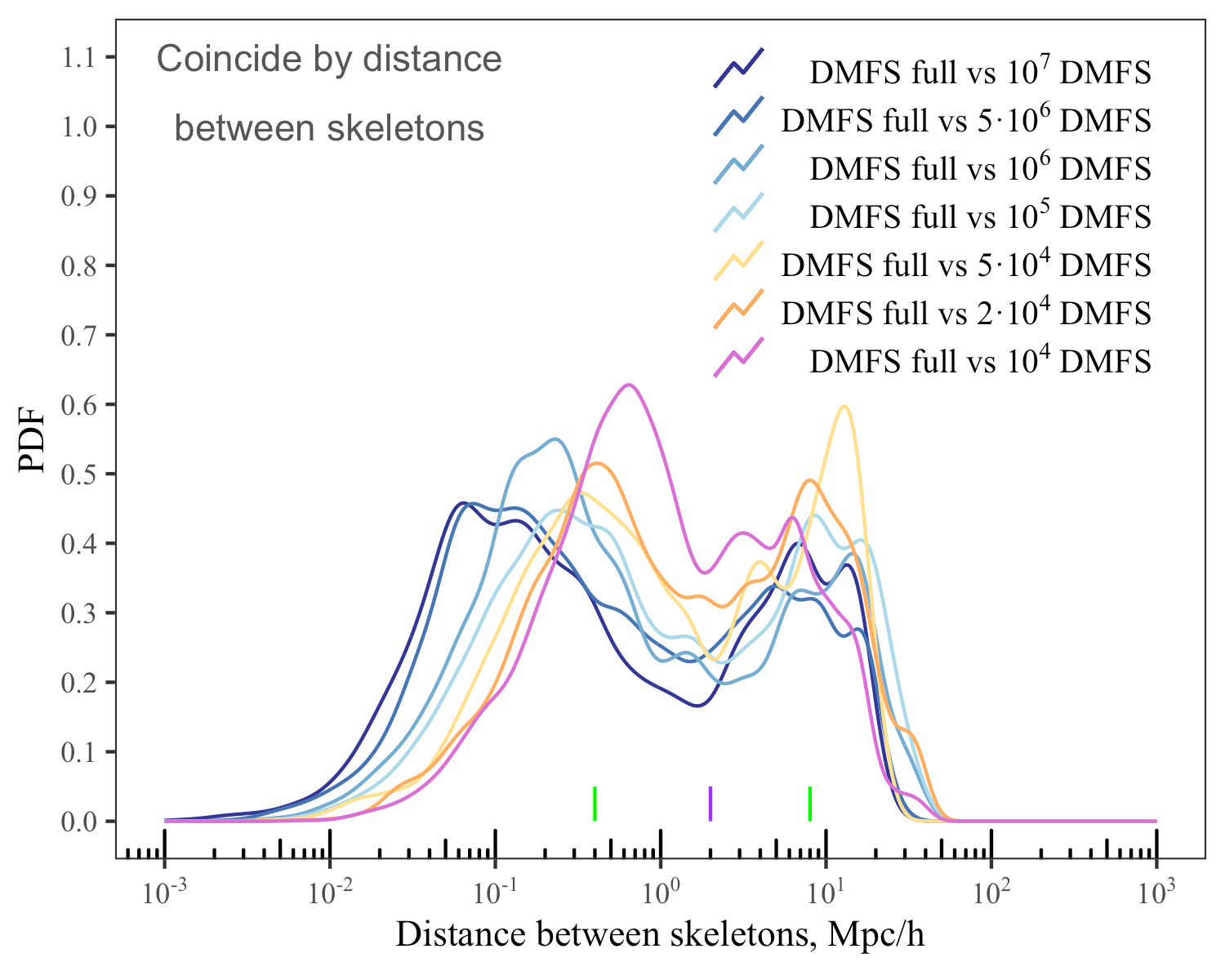}
       \caption{Coincidence by distance between skeletons: probability distribution function (PDF) of the distance of all segments of the DMFS-subsampled from those of the DMFS. The bottom ticks represent the peaks shown in  Fig.~\ref{fig:coin_by_dist}. }
       \label{fig:appendix:subsampling_coincidence}
    \end{figure}

       By visually comparing the different extractions, we find that in all the cases the main features of the filament structure are captured, even though some differences emerge. We attribute some of these to the changes in topology due to the decrease in the number of particles used to obtain the FS. To quantify the differences, we estimate the coincidence between skeletons by distance, as it was done in Sec.\ref{sec:coin_by_dist}. We use the original cube as reference and plot the distribution of the distance between the different skeletons in  Fig.~\ref{fig:appendix:subsampling_coincidence}. If the entire original skeleton was fully recovered, we would obtain an unimodal peaked distribution, whose width would represent minor discrepancies for the position of the skeleton. In contrast, the distribution is bimodal, highlighting how some of the DM filaments do not have sufficient density, when decreaseing the number of particles, to be detected. We also note the shift of the fist peak to right with decreasing  the number of particles in the cube. It means that the samples with a lower number of particles reproduce the filament axis less accurately. The typical distance between the coinciding filaments (the position of the first peak) of the full DMFS and DMFS extracted using  $10^7$ particles differs by 0.1 Mpc/h while the difference between the full DMFS and $10^4$ DMFS is 0.5 Mpc/h. All the other peaks are bracketed between these two values.

        \par
\begin{figure}
        \centering
        \includegraphics[width=1\linewidth]{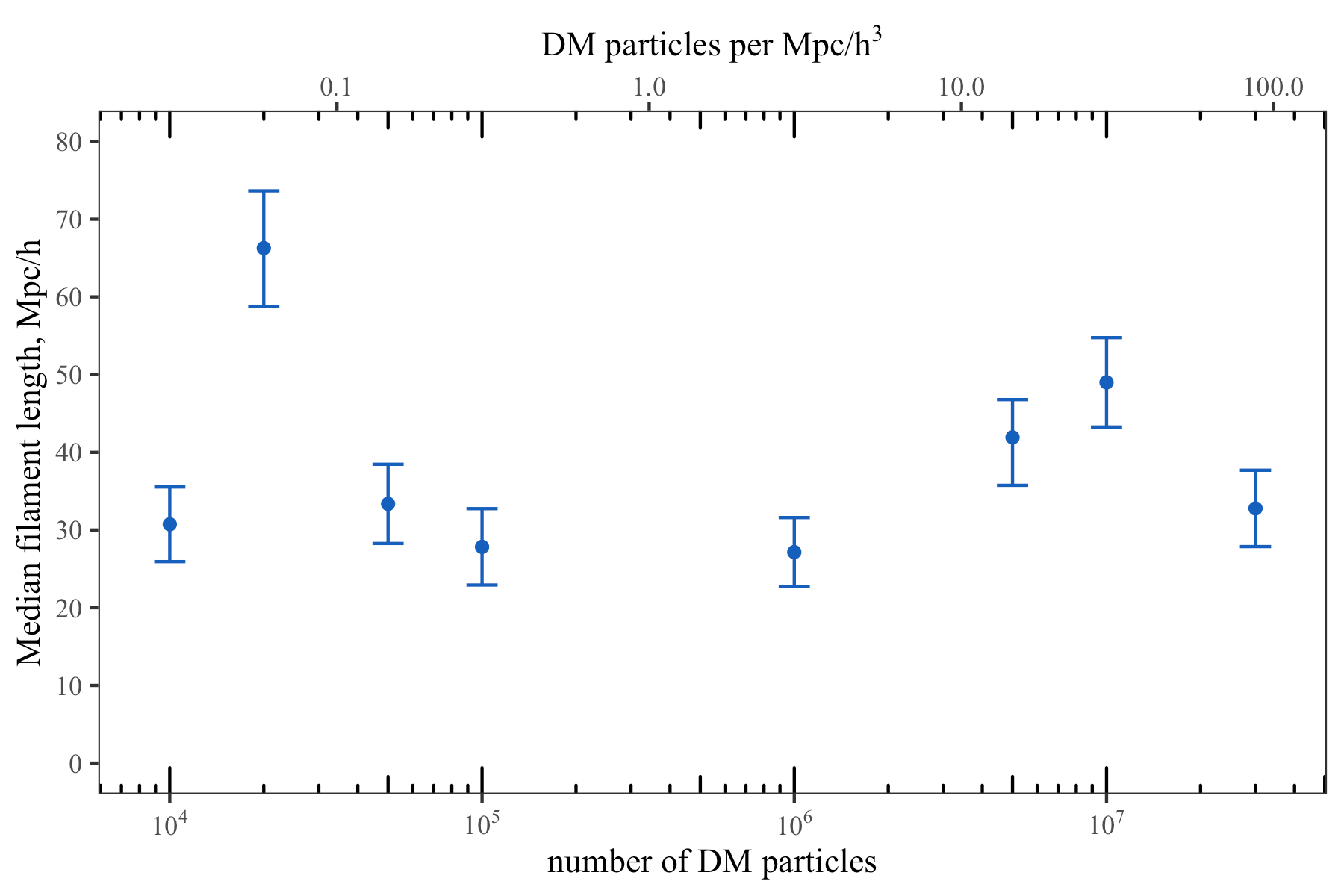}
       \caption{The relation between the median length of single filament and the number of particles used for extraction of skeleton (lower x-axis) and the numebr density of the particles in each cube (upper x-axis). The errorbars show the variance of logonormal distribution of single filament length. }
       \label{fig:appendix:numberofsamples_vs_medianfilength}
    \end{figure}

        We find that, using a different number of particles does not affect the average number of filaments detected: the number varies from 10 to 19, with a mean value of 15.
        We also measured the median filament length in each cube, and contrasted it to the number of DM particles in each cube (or, likewise, to the number density of the particles in each cube) in Fig.~\ref{fig:appendix:numberofsamples_vs_medianfilength}. Overall, the relation is rather flat, suggesting that the subsampling  does not have an effect on the single filament length.  
    
\bsp	
\label{lastpage}
\end{document}